\documentclass[english,aps,prl,twocolumn,superscriptaddress,showpacs]{revtex4}
\usepackage[T1]{fontenc}
\usepackage[utf8]{inputenc}
\usepackage{babel}
\usepackage{amstext}
\usepackage{amsmath}
\usepackage{graphicx}

\usepackage[resetlabels]{multibib}
\newcites{main,suppl}{main,suppl}
\usepackage{placeins}
\usepackage[unicode=true,pdfusetitle,bookmarks=true,bookmarksnumbered=false,bookmarksopen=false,breaklinks=false,pdfborder={0 0 1},backref=false,colorlinks=false]{hyperref}

\newcommand{\angstrom}{\mbox{\normalfont\AA}}
\newcommand{\ignore}[1]{}

\renewcommand{\Im}{Im}
\DeclareMathOperator{\Tr}{Tr}

\begin{document}

%%%%%%%%%%%%%%%%%%%%%
%% Main manuscript (here we use \citemain) %%
%%%%%%%%%%%%%%%%%%%%%

%Title
\title{Electronic Transport in Graphene with Aggregated Hydrogen Adatoms}

%Authors and affiliations
\author{Fernando~Gargiulo}
\affiliation{Institute of Theoretical Physics, Ecole Polytechnique Fédérale de Lausanne (EPFL), CH-1015 Lausanne, Switzerland}
\author{Gabriel~Aut\`es}
\affiliation{Institute of Theoretical Physics, Ecole Polytechnique Fédérale de Lausanne (EPFL), CH-1015 Lausanne, Switzerland}
\author{Naunidh~Virk}
\affiliation{Institute of Theoretical Physics, Ecole Polytechnique Fédérale de Lausanne (EPFL), CH-1015 Lausanne, Switzerland}
\author{Stefan~Barthel}
\affiliation{Institut für Theoretische Physik, Universität Bremen, Otto-Hahn-Allee 1, D-28359 Bremen, Germany}
\affiliation{Bremen Center for Computational Materials Science, Am Fallturm 1a, D-28359 Bremen, Germany}
\author{Malte~R\"osner}
\affiliation{Institut für Theoretische Physik, Universität Bremen, Otto-Hahn-Allee 1, D-28359 Bremen, Germany}
\affiliation{Bremen Center for Computational Materials Science, Am Fallturm 1a, D-28359 Bremen, Germany}
\author{Lisa~R.~M.~Toller}
\affiliation{Institute of Theoretical Physics, Ecole Polytechnique Fédérale de Lausanne (EPFL), CH-1015 Lausanne, Switzerland}
\author{Tim~O.~Wehling}
\affiliation{Institut für Theoretische Physik, Universität Bremen, Otto-Hahn-Allee 1, D-28359 Bremen, Germany}
\affiliation{Bremen Center for Computational Materials Science, Am Fallturm 1a, D-28359 Bremen, Germany}
\author{Oleg~V.~Yazyev}
\affiliation{Institute of Theoretical Physics, Ecole Polytechnique Fédérale de Lausanne (EPFL), CH-1015 Lausanne, Switzerland}

\date{\today}

\begin{abstract}
Hydrogen adatoms and other species covalently bound to graphene act as resonant scattering centers affecting the electronic transport properties and inducing Anderson localization. We show that attractive interactions between adatoms on graphene and their diffusion mobility strongly modify the spatial distribution, thus fully eliminating isolated adatoms and increasing the population of larger size adatom aggregates. Our scaling analysis shows that such aggregation of adatoms increases conductance by up to several orders of magnitude and results in significant extension of the Anderson localization length in the strong localization regime. We introduce a simple definition of the effective adatom concentration $x^\star$, which describes the transport properties of both random and correlated distributions of hydrogen adatoms on graphene across a broad range of concentrations.
\end{abstract}


\maketitle

%Introduction 1
Graphene has unveiled a plethora of unconventional 
transport phenomena \citemain{geim_rise_2007,castro_neto_electronic_2009,transport_review_graphene_2011}, such as the universal minimal conductivity \citemain{tworzydlo_sub-poissonian_2006}, Klein tunneling \citemain{katsnelson_chiral_2006} and the anomalous quantum Hall effect \citemain{zhang_experimental_2005,novoselov_two-dimensional_2005}. On the applied side, graphene is interesting because of its exceptionally high charge-carrier mobility, which is typically limited by the presence of
various types of disorder. Resonant scattering impurities, such as
chemical functionalization defects \citemain{wehling_resonant_2010} and dislocations \citemain{gargiulo_topological_2013}, show
the most pronounced effects on charge-carrier transport in graphene. Hydrogen adatoms represent a prototypical resonant 
scattering impurity, which can be experimentally introduced in a
controlled fashion \citemain{ni_resonant_2010} and allows for a simple theoretical description \citemain{wehling_local_2007}. 
A hydrogen adatom covalently binds to a single carbon atom of graphene resulting in rehybridization into the $sp^3$ state, thus effectively removing that site from the honeycomb network of $p_z$ orbitals.
This gives rise to a zero-energy state localized around the defect, and results in the resonant scattering of charge carriers. 
%and resonant backscattering for low angle grain boundaries \citemain{gargiulo_topological_2013}.
 
At a fundamental level, the classical scaling theory of Anderson transition predicts complete localization of the electronic spectrum in two dimensions (2D), regardless of the amount of disorder \citemain{kramer_localization:_1993}. For hydrogenated graphene, a model based on massless Dirac fermions with $\delta$-function-like potentials confirms this prediction of the unitary class,
% that the entire spectrum is localized, 
though in 2D systems localization lengths can be strongly energy-dependent and, eventually, very large \citemain{evers_anderson_2008}. However, no unanimous consensus has been reached since experiments on hydrogenated graphene point towards metal-insulator transition, theoretically justified by the presence of electron-hole puddles (2D percolation class) \citemain{bostwick_quasiparticle_2009,song_effects_2011,jayasingha_situ_2013,adam_density_2008}.

Early works treating finite concentrations of resonant impurities in graphene assumed that the total scattering cross-section deviates little from the incoherent addition of the individual cross-sections, for example in the Boltzmann equation framework  \citemain{shon_quantum_1998}. This picture is valid for low defect concentrations, low charge-carrier densities and random adatom distributions. A better description requires including the effect of coherent superposition of wavefunctions scattered by distinct adatoms \citemain{wehling_resonant_2010,wehling_impurities_2009,ferreira_unified_2011,Cresti2013}.  This is particularly important when impurities are in proximity to each other, with the limiting case being the formation of compact clusters in which hydrogen adatoms populate neighboring carbon atoms \citemain{palacios_vacancy-induced_2008,leconte_magnetism-dependent_2011,trambly_de_laissardiere_conductivity_2013}.  
Indeed, the overall short-range attractive interaction 
between individual hydrogen adatoms on graphene \citemain{boukhvalov_hydrogen_2008,lin_hydrogen_2008}, combined with their relatively 
high diffusion mobility at room temperature \citemain{Yazyev2008,herrero_vibrational_2009,moaied_theoretical_2014},
suggests a high degree of spatial correlation between adatoms.
 
In this Letter, we address the effects of spatial correlation of resonant impurities on electronic transport in graphene.
The equilibrium configurations of hydrogen adatoms on graphene,  obtained by means of Monte-Carlo simulations, show a strong tendency towards aggregation into small clusters essentially eliminating isolated adatoms. Electronic transport properties investigated using the Landauer-B\"uttiker approach complemented with
kernel polynomial method calculations show that aggregation dramatically increases both the conductivity and the localization length. We propose a unified framework to account for the effects of spatial correlation of resonant scattering centers on  electronic transport in graphene.

%Brief sketch of the methods

\begin{figure}
\includegraphics[width=8.2cm]{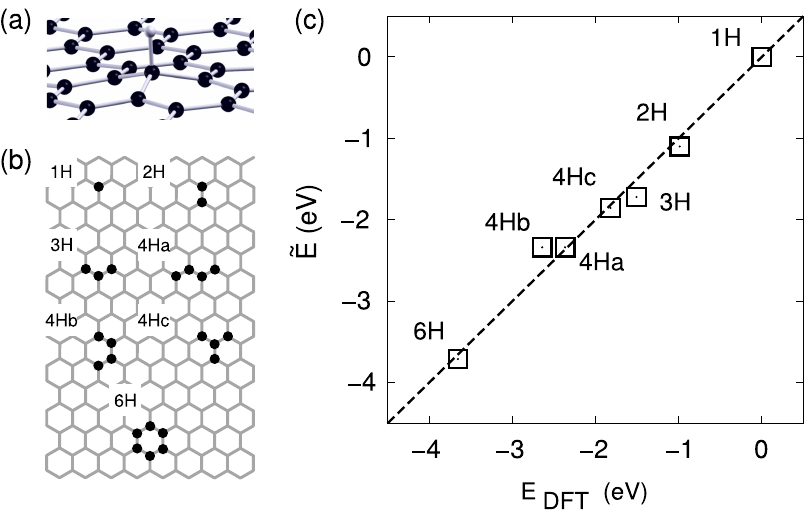}
\caption{(a) Atomic structure of a hydrogen adatom covalently bound to graphene. (b) Structures of small clusters of hydrogen adatoms used for fitting the pair potential of Eq.~(\ref{cluster_expansion}) (c) Predicted energy $\tilde{E}$ of aggregation of hydrogen adatoms as a function of aggregation energy $E_{\textrm{DFT}}$ calculated from first principles for the set of adatom clusters shown in panel (b).}
\label{figure1_DFT}
\end{figure}

Upon adsorption, a hydrogen adatom covalently 
binds to a single carbon atom of graphene changing its hybridization state to 
$sp^3$ and its coordination sphere to tetrahedral as shown in Fig.~\ref{figure1_DFT}(a) \citemain{Yazyev2007}. 
The covalent binding of a second adatom to the nearest neighbor 
carbon atom partially releases the elastic energy due to the change of coordination sphere, thus resulting in effective attractive interaction between adatoms \citemain{boukhvalov_hydrogen_2008,lin_hydrogen_2008}.
This suggests that the interaction between adatoms can be accurately
described using a short-ranged pair potential. In our study, we 
expand an Ising-like interaction energy $E$ up to the second nearest-neighbor term 
\begin{equation}
 E=\gamma_1 \sum_{\langle i,j \rangle} s_i s_j + \gamma_2 \sum_{\langle \langle i,j \rangle \rangle} s_i s_j ,
  \label{cluster_expansion}
\end{equation}
where $\gamma_1$ and $\gamma_2$ are the corresponding first and second nearest-neighbor parameters. Here, $s_i = 1$ if a carbon atom $i$ is populated by an adatom, otherwise $s_i = 0$.
Parameters $\gamma_1$ and $\gamma_2$ are obtained by fitting the 
interaction energies of adatoms calculated from first principles \citemain{suppl} 
for a set of small adatom aggregates shown in Fig.~\ref{figure1_DFT}(b). Under the assumption of single-side functionalized graphene, the 
obtained parameters $\gamma_1=-1.182$~eV and $\gamma_2=0.484$~eV \citemain{note_both_sides} signify a considerable first-nearest-neighbor
attraction alongside a weaker second-nearest-neighbor repulsion.
The excellent agreement between the interaction energy $\tilde{E}$, estimated using the fitted $\gamma_1$ and $\gamma_2$, and the first-principles values $E_{\rm DFT}$, confirms the applicability
of the short-range pair potential form (\ref{cluster_expansion}) for describing small clusters [Fig.~\ref{figure1_DFT}(c)].

%Figure 1 - DFT

\begin{figure}[b]
\includegraphics[width=8.2cm]{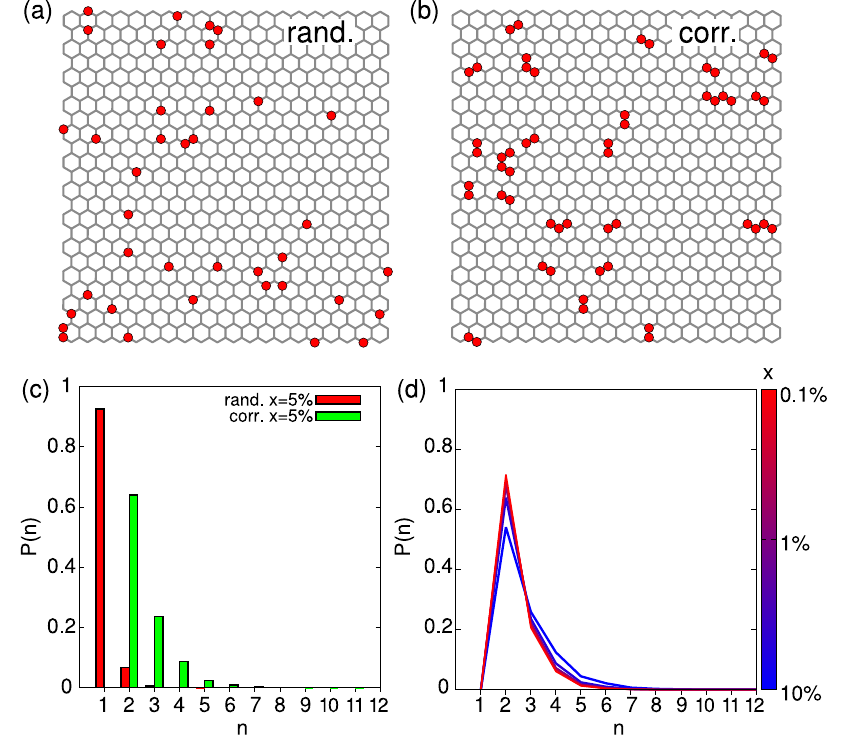}
\caption{(Color online) Representative configurations of (a) randomly distributed and (b) correlated hydrogen adatoms on graphene at $x=5\%$ concentration. (c) Comparison of random and correlated adatoms cluster size distributions $P(x)$ at $x=5\%$ concentration. 
(d) Cluster size distributions $P(x)$ of correlated hydrogen adatoms at different concentrations. All correlated configurations are obtained by means of Monte-Carlo simulations carried out at $T=300$~K.}
\label{figure2_MC}
\end{figure}

In order to assess the effect of interaction between adatoms on their spatial distribution and transport properties we perform Monte-Carlo simulations using the introduced pair potential (\ref{cluster_expansion}).
The simulations employ a Monte-Carlo move based on the displacement of a randomly selected adatom to a random unoccupied carbon atom in combination with the Metropolis acceptance criterion \citemain{metropolis_equation_1953}. We considered models containing up to
$N_{\rm C}=10^{6}$ carbon atoms ($165 \times 165$~nm$^2$) and adatom concentrations
$x=N_{\rm H}/N_{\rm C}$ ranging from 0.1\% to 10\%.  
All simulations have been performed at $T=300K$. A representative configuration 
of randomly distributed adatoms [Fig.~\ref{figure2_MC}(a)] is 
compared with a configuration obtained from a Monte-Carlo simulation [Fig.~\ref{figure2_MC}(b)] at $x=5\%$. Further details are revealed
by comparing the cluster size distributions $P(n)$ [Fig.~\ref{figure2_MC}(c)], with adatoms populating neighboring carbon atoms being assigned to the same cluster. 
In the case of a random distribution most adatoms are isolated
($n = 1$), while the occurrence of clusters ($n > 1$) is merely a probabilistic effect. 
In contrast, no isolated adatoms are found in the presence of interactions, with the most abundant species being adatom dimers ($n = 2$). The size 
distribution for the correlated case shows a longer tail with a significant probability of observing up to $n=6$ clusters. 
The dependence of $P(n)$ on adatom concentration $x$ is relatively 
weak [Fig.~\ref{figure2_MC}(d)].

\begin{figure}
\includegraphics[width=8.2cm]{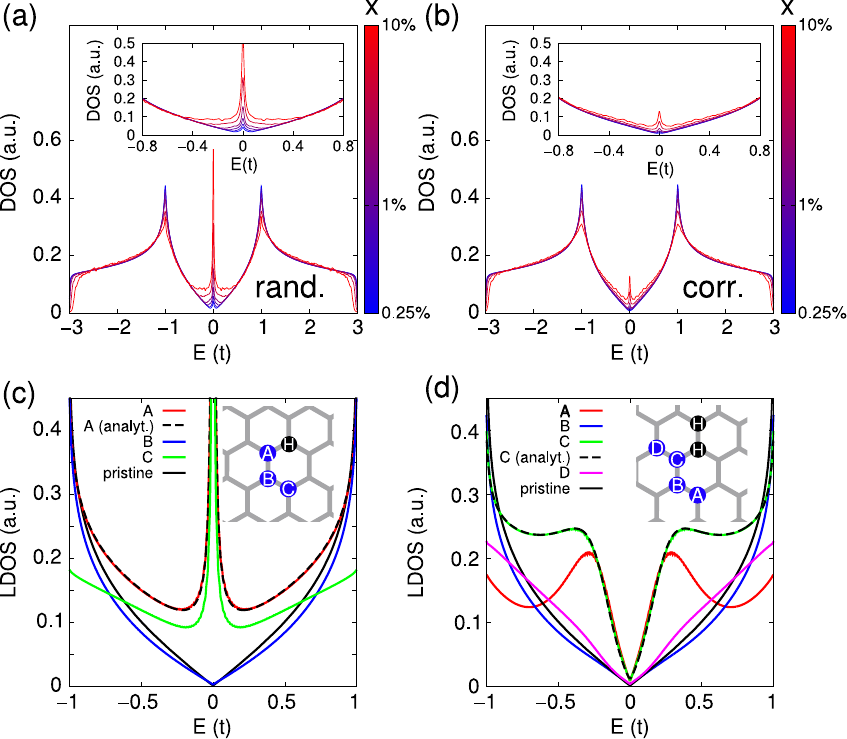}
\caption[fake]{(Color online) Density of states of graphene in the presence of (a) randomly distributed and (b) correlated hydrogen adatoms at different concentrations. Local density of states (LDOS) on carbon atoms in the vicinity of (c) an isolated hydrogen adatom and (d) a dimer of hydrogen adatoms. In panels (c,d) LDOS referred to as ``analyt.'' have been obtained using the analytical Green's function calculations \citemain{fernando_gabriel}.} 
\label{figure3_DOS}
\end{figure}

%DOS discussion - first comments
We now focus on electronic and transport properties calculated using the nearest-neighbor tight-binding Hamiltonian for $p_{z}$ orbitals 
\begin{equation}
H=-t\sum_{\langle i,j\rangle}[c_{i}^{\dagger}c_{j}+{\rm h.c.}]
 \label{TB}
\end{equation}
with the hopping integral constant $t=2.7$~eV \citemain{castro_neto_electronic_2009}.
An adsorbed hydrogen atom is modeled by excluding the $p_{z}$ orbital of the carbon atom to which it is bound as a consequence of $sp^3$ hybridization, making it similar to a vacancy defect \citemain{Pereira2006,Yazyev2007,Yazyev2008}. We stress that adatoms do not introduce coupling between the sites owing to the same sublattice of the bipartite lattice of graphene, thus maintaining electron-hole symmetry of the electronic spectrum.
Figures~\ref{figure3_DOS}(a,b) compare the density of states (DOS) 
of graphene with random and correlated distributions of adatoms at different concentrations. In the case of a random distribution one observes a
strong peak at $E=0$ due to the resonant modes originating from isolated adatoms \citemain{Pereira2006,wehling_impurities_2009}. The corresponding wave function is localized on the sublattice opposite to that of the carbon atom binding the adatom and decays from the defect position \citemain{basko_resonant_2008}. 
At high adatom concentrations, $x >1\%$, the $E=0$ peak is accompanied by flat density regions at higher energies with a noticeable overall renormalization of the DOS, in agreement with previous calculations \citemain{Pereira2006}. In comparison, the DOS calculated for the correlated impurity configurations shows a less intense peak at $E=0$ and an increased weight for $-0.9t < E <0.9t$ that is more evident at higher concentrations. This change is a direct consequence of different cluster size distributions. The dominant cluster type in the 
case of the correlated adatom distribution is the dimer ($n = 2$), which is known 
to be non-resonant \citemain{wehling_local_2007,fernando_gabriel,trambly_de_laissardiere_conductivity_2013},
meaning that no localized states emerge at any energy. The local density of states (LDOS) calculated on neighboring atoms of an isolated single adatom shows a singularity at $E=0$ [Fig.~\ref{figure3_DOS}(c)]. In contrast, an enhancement 
of the LDOS in a broad energy region $-t < E < t$ is observed on certain carbon atoms in the vicinity of the adatom dimer [Fig.~\ref{figure3_DOS}(d)]. 
The numerical LDOS for selected impurities compared to the results of analytical Green's function calculations shows no discrepancies \citemain{fernando_gabriel}. 

%%%%%%%%%%%%%%%%
%N. B. Since we include a citation in the caption we need to add a short caption. Otherwise it clashes with multibib.
%%%%%%%%%%%%%%%%
%Figure 3 - DOS

Based on these observations we conclude that the residual peak at $E =0$ in the case of the correlated distribution is due to the $n > 2$ adatom aggregates, which break the bipartite symmetry of graphene (that is, populate different number of sites in the two sublattices). This gives rise to resonant modes at $E=0$ akin to isolated adatoms. All odd-$n$ aggregates and certain configurations of even-$n$ clusters lead to resonant modes at $E =0$. Judging on the cluster size distributions [Fig.~\ref{figure2_MC}(d)], the largest contribution comes from adatom trimers. This suggests that
adatom aggregation has strong effects on the electronic transport properties
that we investigate by performing a scaling analysis of conductivity $g$ using the Landauer-B\"{u}ttiker approach \citemain{buttiker_generalized_1985}. In this approach,
the conductance $G(E)$ is given as $G(E)=G_{0}T(E)$, where 
$G_0$ is the conductance quantum and $T(E)$ is the transmission probability across the scattering region at energy $E$. We assume a two-terminal device configuration with a scattering region of width $W=40$~nm perpendicular to the current direction, and of variable 
length 1~nm~$< L <$~60~nm. The scattering region is attached to pristine graphene contacts and populated by 
adatoms according to concentrations and statistical distributions 
discussed above. Further details of our methodology are given in the Supplementary Material \citemain{suppl}.

\begin{figure}
\includegraphics[width=8.2cm]{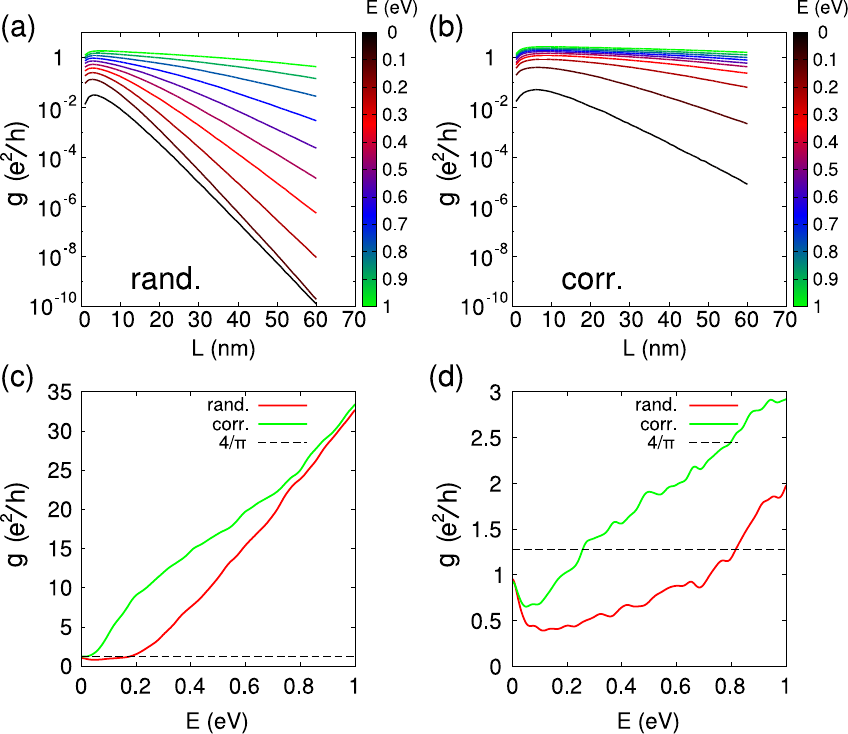}
\caption{(Color online) (a,b) Averaged conductivity $g_{\rm typ}$ for the random and correlated adatoms distributions, respectively, as a function of scattering region length $L$ at different energies $E$ and at $x = 5\%$ concentration. (c,d) Conductivity $g$ as a function of energy $E$ calculated using the kernel polynomial method for both adatom distributions at concentration $x=0.5\%$ and $x=5.0\%$, respectively. The dashed lines show the minimum conductivity $g = 4/\pi$.}
\label{figure4_G}
\end{figure}

The characteristic functional laws for the conductivity $g=G \times L/W$ in the ballistic, diffusive and localized transport regimes are $g \propto L$, $g = const$ and $g \propto \exp(-L/\xi_{\rm loc})$, respectively, where $\xi_{\rm loc}$ is the localization length.
In the localized regime $ln(g)$ follows a broad positively-skewed distribution, which means that $g$ can show strong fluctuations depending on the exact configuration of defects, especially in the presence of strong localization \citemain{choe_effect_2012}. An estimate of the mean value for such a distribution is given by the typical conductivity $g_{\rm typ}=\exp\langle \ln(g)\rangle$ \citemain{choe_effect_2012,uppstu_obtaining_2014}. In our scaling analysis, $g_{\rm typ}$ has been obtained averaging over 9600 disorder realizations.
Figure~\ref{figure4_G}(a,b) shows $g_{\rm typ}$ as a function of scattering region length $L$ at different energies for the random and correlated impurity distributions, both at $x=5\%$ concentration. We observe a short transition from ballistic to diffusive and subsequently to localized regime within the first 10~nm. The crossover lengths are expected to be of the order of the elastic mean free path $\xi_{\rm el}$ and localization length $\xi_{\rm loc}$, respectively. The general trend is that the localized regime is accentuated at low energy, whereas at higher energy the onset of exponential decay occurs at larger $L$ and the absolute slope of the conductance curves is smaller. The scaling of $g$ also depends strongly on the impurity concentration. For low adatom concentrations ($x \lesssim 0.5\%$) the onset of the localized regime is only observable in the vicinity of the Dirac energy (see Supplementary Material \citemain{suppl} for complete results).

\begin{figure}[t]
\includegraphics[width=8.2cm]{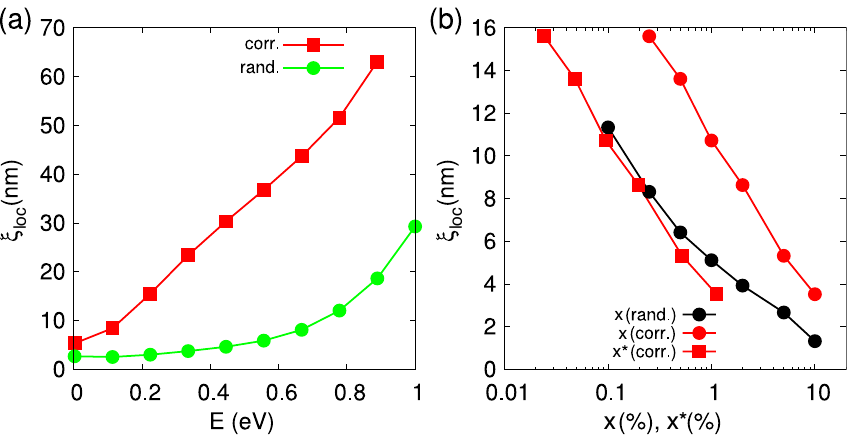}
\caption{(Color online) (a) Localization length $\xi_{\rm loc}$ as a function of charge-carrier energy $E$ for the case of random and correlated adatom distribution at $x=5\%$.
(b) Charge-carrier localization length $\xi_{\rm loc}$ at low energy ($E= 10^{-3}t = 2.7$~meV) as a function of simple concentration $x$ and  effective concentration $x^\star$ for the random and correlated distributions.}
\label{figure5_loc}
\end{figure}

We stress that the conductance curves vary smoothly in the whole range of $E$ and $x$, never showing singularities which would indicate a phase transition such as the metal-insulator transition (MIT). Thus, we ascribe the non-observance of the localized regime to an insufficient scattering region length of our model, which is shorter than $\xi_{\rm loc}$ for some choices of $E$ and $x$. From Figure~\ref{figure4_G}(a,b) it follows that the presence of spatial correlation between adatoms enhances the conductance by up to five orders of magnitude in the vicinity of the Dirac point ($E=2.7$~meV). This is a direct consequence of the suppressed weight of low-energy resonant states, as explained above. A closely related effect is a significant increase of the localization length $\xi_{\rm loc}$ at all energies upon adatom aggregation. $\xi_{\rm loc}$ was obtained by fitting conductance curves to the expected law $g_{\rm typ}\propto \exp({-{L}/{\xi_{\rm loc})}}$ \citemain{lee_disordered_1985}, and is shown in Fig.~\ref{figure5_loc}(a).
At $x=5\%$, the localization length is well defined for the entire range of investigated energies $-1$~eV~$< E <$~1~eV, whereas at lower concentrations it is well defined only in proximity of the Dirac point (see Supplementary Material \citemain{suppl}). However, as long as a finite positive $\xi_{\rm loc}$ can be determined, it proves to be up to an order of magnitude larger for the correlated adatom distribution compared to the random case.

%Figure 4 - Conductivity

Our Landauer-B\"{u}ttiker results are complemented by the Kubo formula calculations based on the kernel polynomial method (KPM) \citemain{weise_kernel_2006}. This approach allows for a direct calculation of the DOS and conductivity independent of sample geometry and contacts. Figure~\ref{figure4_G}(c,d) shows the conductivity $g$ for adatom concentrations $x=0.5\%$ and $x=5\%$, and both random and correlated adatom distributions. In comparison to the Landauer-B\"{u}ttiker ansatz, the results are similar at both concentrations, confirming the enhancement of conductivity upon aggregation of adatoms. 
The only difference appears when the charge-carrier energy $E$ falls into the resonant peak region where the KPM conductivity exceeds the Landauer-B\"{u}ttiker results. This can be proved to be an effect of the increased DOS due to the formation of an impurity band. While the KPM conductivity reflects the excess of states at the Fermi energy, in the Landauer-B\"{u}ttiker ansatz the number of charge carriers is limited by the DOS of the pristine graphene leads, and subsequently the conductivity is determined only by the mean free path and the localization length of the scattering region. We note that upon doping the leads, the LB conductivity at low energies increases while the localization length is unaffected (see Supplementary Material \citemain{suppl}).

%Figure 5 - Localization length and effective concentration

%Focus on localization - Effective concentration
On the basis of the identification of resonant adatom clusters we introduce an effective concentration 
\begin{equation}
x^\star = \frac{1}{N_{\mathrm{C}}} \sum_i N_i |n_{i}^{\mathrm{A}}-n_{i}^{\mathrm{B}}|, 
 \label{xstar}
\end{equation}
where $N_i$ is the number of instances of the cluster configuration $i$, and $n_{i}^{\mathrm{A}}$ ($n_{i}^{\mathrm{B}}$) is the number adatoms bound to carbon atoms in sublattice A(B) in this configuration. The main contribution to $x^{\star}$ comes primarily from adatom trimers as discussed above. 
We assume that the contribution of non-resonant clusters to the total scattering cross-section can be neglected. It should be stressed that this approach accounts for the diverse scattering effect of different clusters. In order to validate the applicability of effective concentration $x^\star$, we compare $\xi_{\rm loc}(x)$ with $\xi_{\rm loc}(x^\star)$ at low charge-carrier energy for the correlated impurity case [Fig.~\ref{figure5_loc}(b)].  
One can see that replacing $x$ with effective concentration $x^\star$
brings $\xi_{\rm loc}(x^\star)$ evaluated for the case of the correlated impurity distribution in good agreement with $\xi_{\rm loc}(x)$ calculated for the random distribution.   
The agreement is particularly good in the low concentration regime where randomly distributed impurities consist almost exclusively of isolated monomers. Deviations at higher adatom concentrations can be ascribed to the inter-cluster interference effects, which become important at the reduced average distance between the clusters.   

%Final discussion - Conclusions and sum up
To summarize, hydrogen adsorbed on graphene has a strong tendency toward aggregation, resulting in the formation of small clusters and fully eliminating isolated adatoms. Some of the larger clusters, notably trimers, are 
responsible for the residual resonant scattering, but the overall 
conductance and localization length dramatically increases upon 
aggregation.
Within the range of parameters investigated in our work, we find no metal-insulator transition, with the graphene spectrum being fully localized. The predicted effects of adatom aggregation can be investigated experimentally by varying the temperature regimes, since the diffusion 
of hydrogen adatoms occurring at normal conditions can be effectively suppressed at low temperatures. Alternatively, time-resolved transport measurements should evince a rise in conductivity upon formation of adatom clusters.

%Acknowledgements
We would like to thank C.~W.~J.~Beenakker, P.~Lugan and V.~Savona for discussions. This work was supported by the Swiss National Science Foundation (Grant No. PP00P2\_133552). 

\FloatBarrier

%Here we put the main bibliography
\bibliographystylemain{apsrev4-1}
\bibliographymain{main}

%%%%%%%%%%%%%%%%%%%%%%%%%%%%%%%%%%%%%%%%%%%%%%%%%%%%
%% Supplementary material (here we use \citecuppl)%%
%%%%%%%%%%%%%%%%%%%%%%%%%%%%%%%%%%%%%%%%%%%%%%%%%%%%

%Reset counters
\setcounter{equation}{0}
\setcounter{figure}{0}
\setcounter{table}{0}
\setcounter{page}{1}
\makeatletter

%New command for figures and references enumeration starting with S
\renewcommand{\theequation}{S\arabic{equation}}
\renewcommand{\thefigure}{S\arabic{figure}}
\renewcommand{\bibnumfmt}[1]{[S#1]}
\renewcommand{\citenumfont}[1]{S#1}

% Force a newpage
\clearpage

%Force one column format
\widetext

%Manual format of the title
\begin{center}
\textbf{\large Supplementary Material for \\
``Electronic Transport in Graphene with Aggregated Hydrogen Adatoms''}
\end{center}

\subsection{First-principles calculations of the energies of hydrogen adatom clusters}

First-principles calculations of the interaction energies of hydrogen adatoms on graphene have been performed within the density functional theory (DFT) framework employing the generalized gradient approximation (GGA) to the exchange-correlation functional \citesuppl{s_perdew_generalized_1996}. Ultrasoft pseudopotentials \citesuppl{s_vanderbilt_soft_1990} for carbon and hydrogen atoms have been used in combination with a plane-wave
basis set with a kinetic energy cutoff of 30~Ry for the wavefunctions. Models of hydrogen adatom clusters are based on a graphene $6\times6$ supercell with 15~\AA\ of vacuum separating the periodic replicas. We used a $2\times2\times1$ Monkhorst-Pack k-point mesh for the Brillouin zone integration \citesuppl{s_monkhorst_special_1976}. All hydrogen adatom cluster models were relaxed until a maximum force of 0.15~eV/\AA~on individual atoms was reached. We verified that the chosen parameters provide sufficiently accurate total energies. All calculations have been performed using the {\sc pwscf} code of the 
{\sc Quantum ESPRESSO} package \citesuppl{s_giannozzi_quantum_2009}.

The interaction energy of a cluster of hydrogen adatoms calculated from first principles $E_{\rm DFT}$ is defined as
\begin{equation}
E_{\rm DFT}=E_{{\rm gr}+n{\rm H}}-E_{\rm gr}-n\left(E_{{\rm gr}+{\rm H}}-E_{\rm gr}\right),
 \label{Energy_DFf}
\end{equation}
where $E_{{\rm gr}+n{\rm H}}$, $E_{\rm gr}$ and $E_{{\rm gr}+{\rm H}}$ are the total energies of graphene with a cluster of $n$ hydrogen adatoms, pristine graphene and graphene with a single hydrogen adatom, respectively.

\begin{figure}[b]
\includegraphics[width=12cm]{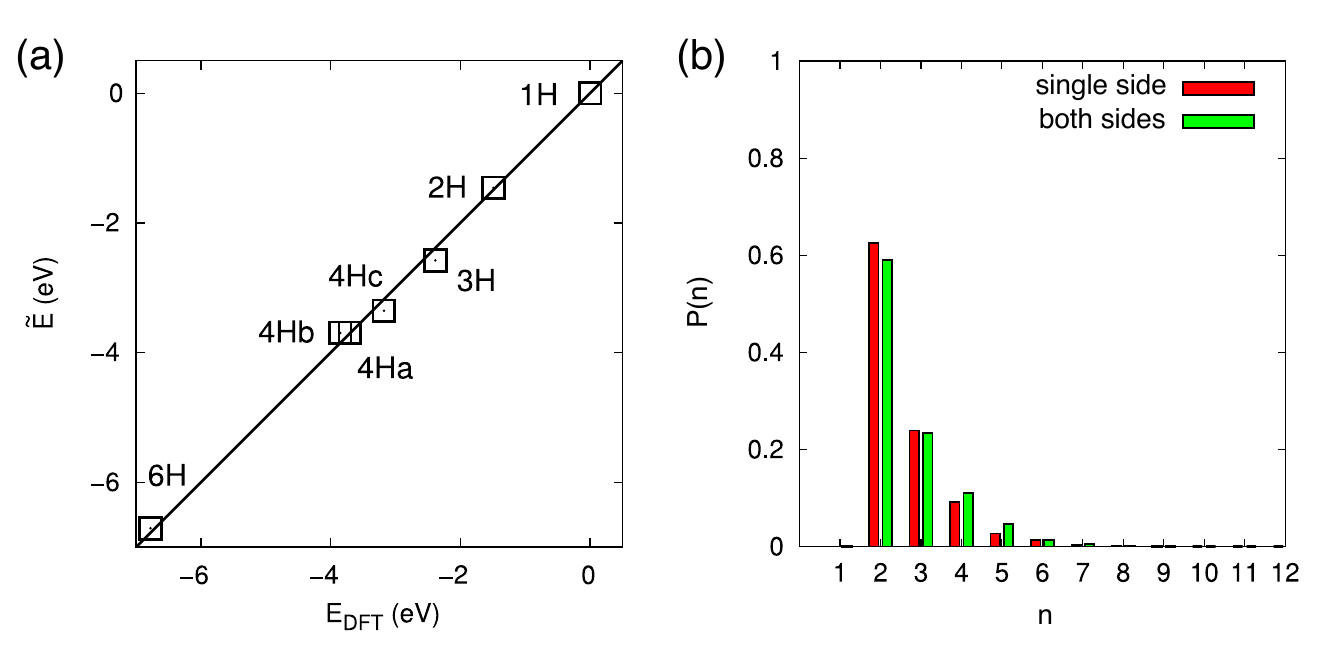}
\caption{(Color online) (a) Predicted energy $\tilde{E}$ of aggregation of hydrogen adatoms as a function of aggregation energy $E_{\textrm{DFT}}$ calculated from first principles for the set of small clusters shown in Fig.~1(b) of the main text with adatoms adsorbed on both sides of graphene.
(b) Comparison of the cluster size distributions $P(n)$ for the 
cases of single-side and both-sides adatom adsorption at $x=5\%$ concentration and $T=300$~K.}
\label{FigS1}
\end{figure}

In the main text of our manuscript we focused on the situation where hydrogen adatoms are deposited on a single side of graphene. The values of the fitted parameters in expr. (1) of the main text are  $\gamma_1=-1.182$~eV and $\gamma_2=0.484$~eV. 
This scenario is relevant to the case of graphene on a substrate, however in the situation of suspended graphene both sides of graphene
are available for binding adatoms. We investigated this situation by studying the same set of adatom clusters as shown in Fig.~1(b) of the main text, but with adatoms placed on the opposite sides of graphene sheet when functionalized carbon atoms belong to different sublattices. The fitted interaction parameters are $\gamma_1 = -1.461$~eV and $\gamma_2 = 0.342$~eV. The excellent agreement between the estimated interaction energies $\tilde{E}$ and the first-principles values $E_{\rm DFT}$ is illustrated in Fig.~\ref{FigS1}(a). This case is characterized by a stronger attractive contribution and a weaker repulsion compared to the single-side adsorption, thus reflecting the known tendency of forming more stable adatom aggregates upon adsorption on both sides \citesuppl{s_lin_hydrogen_2008,s_boukhvalov_hydrogen_2008}.
The cluster size distributions $P(n)$ calculated for single-side and both-sides adsorption at adatom concentration $x=5\%$ and $T=300$~K are compared in Fig.~\ref{FigS1}(b). While the distributions are qualitatively very similar, one notes that both-sides adsorption exhibits a somewhat stronger tendency to form larger clusters.  

\subsection{Monte-Carlo simulations of the hydrogen adatoms aggregation}

We performed Monte-Carlo simulations with an elementary trial move being the displacement of a randomly chosen adatom to a random carbon atom not populated by another adatom. This move insures the fulfillment of detailed balance. The Metropolis algorithm has been employed for the acceptance/rejection criterion. Once a move has been performed, the system is updated from the old configuration $S_{\mathrm{old}}$ to the new one $S_{\mathrm{new}}$ with a probability
\begin{equation}
   P(S_{\mathrm{old}} \rightarrow S_{\mathrm{new}}) = \mathrm{min} (1, e^{-\beta [\tilde{E}(S_{\mathrm{new}})-\tilde{E}(S_{\mathrm{old}})]}),
 \label{Metropolis}
\end{equation}
where $\beta$ is the inverse temperature $1/(k_{\mathrm{B}}T)$. In all our simulations $T=300~\mathrm{K}$. 
The number of equilibration steps $N_{\mathrm{eq}}$ disregarded from statistical sampling varied between 
$10^6$ and $1.6 \times 10^7$, depending on adatom concentration (larger concentrations need more equilibration steps). The total number of steps in our simulations varied between $10^7$ and $3 \times 10^7$.
The equilibration efficacy of our simulation was tested by comparing certain equilibrium properties such as the total energy and cluster size distribution. The properties obtained from Monte-Carlo simulations performed with and without temperature annealing show negligible differences.

\subsection{Landauer-B\"{u}ttiker electronic transport calculations and scaling analysis of conductivity}

\begin{figure}[b]
\includegraphics[width=8cm]{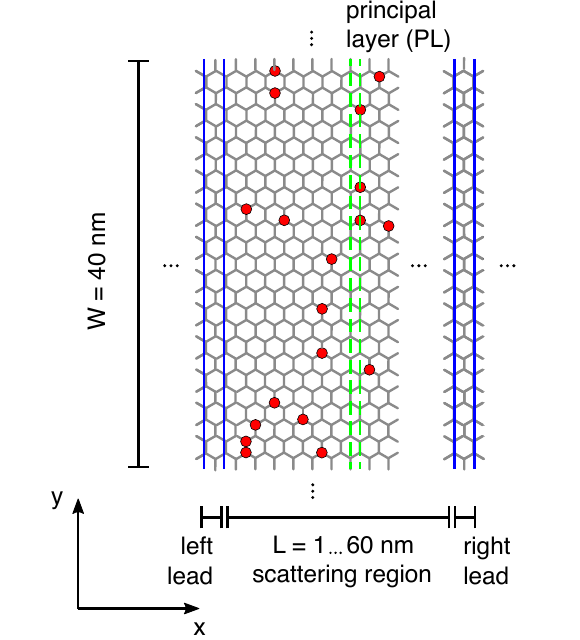}
\caption{(Color online) Schematic drawing of the two-terminal configuration employed for investigating the transport properties of 
graphene with hydrogen adatoms. The transport direction is along the $x$ axis while the system is periodic along the $y$ axis. The periodicity along the $y$ axis is $W=40$~nm. The unit cells of the left and right leads composed of pristine graphene are indicated by blue lines. The scattering region is populated by adatoms either randomly or 
according to the configurations produced by Monte-Carlo simulations in the case of correlated adatom distributions. One of the principal layers (PL) in the scattering region is indicated by means of green dashed lines.
}
\label{FigS2}
\end{figure}

In order to investigate the transport properties of graphene with resonant scattering impurities we perform Landauer-B\"{u}ttiker
calculations in a two-terminal configuration with a scattering region composed of hydrogenated graphene attached to two semi-infinite leads of pristine graphene, as shown in Fig.~\ref{FigS2}. The overall configuration is periodic along the transverse direction $y$. 
For such a setup, conductance as a function of energy $G(E)$ is given by 
\begin{equation}
G(E)=\frac{W}{2\pi}\int_{-\frac{\pi}{W}}^{\frac{\pi}{W}}T(E,k_{\parallel}) dk_{\parallel},
\end{equation}
where $T(E,k_{\parallel})$ is the transmission probability and $k_{\parallel}$ is the momentum along $y$ \citesuppl{s_buttiker_generalized_1985}. Due to the large width $W=40$~nm of the model employed, transmission is only evaluated at the $\Gamma$ point ($k_{\parallel} = 0$). 

In order to calculate $T(E)$ we decompose the scattering region into principal layers (PLs), that is, the layers in which the atoms are coupled at most to those located in the next layer. Since our tight-binding model is limited to first-nearest-neighbor interactions and the transport direction is oriented along zig-zag direction [see Fig.~\ref{FigS2}], the minimal width of the PL is $d_{\rm PL}=\frac{\sqrt{3}}{2}d_{\rm CC}$, where $d_{\rm CC}=1.42~\angstrom$ is the carbon-carbon bond length. The Hamiltonian restricted to the $i$-th principal layer is $H_i$, while $t_i$ is the tight-binding hopping matrix connecting $i$-th and $i+1$-th principal layers. We introduce an imaginary cleavage plane between the $n$-th and $n+1$-th principal layers dividing the system into two independent parts. We define $g_{n}^{\rm L}$ and $g_{n+1}^{\rm R}$ as the surface Green's functions of the two non-interacting semi-infinite systems located on the left and on the right sides of this cleavage plane, respectively.

Following Ref.~\citesuppl{s_mathon_theory_2001}, the transmission is given by    
\begin{equation}
           T(E)=\Tr[{T}_{n} \Im (g^{\mathrm{L}}_{n}){T}_{n}^{\dagger} \Im (g^{\mathrm{R}}_{n+1})],
           \label{Kubo}                                          
\end{equation}
where operator ${T}_{n}$ is defined as
\begin{equation}
 	   {T}_{n}=t_{n}(1-g_{n+1}^{\mathrm{R}} t_{n}^{\dagger} g_{n}^{\mathrm{L}} t_{n} )^{-1}.
\end{equation}
The choice of the position of the cleavage plane is immaterial because of current conservation.

Surface Green's functions $g^{\mathrm{L}}_{n}$ and $g^{\mathrm{R}}_{n}$ can be related to the preceding (successive) surface Green's functions $g_{n-1}^\mathrm{L}$ ($g_{n+1}^\mathrm{R}$) by applying the Dyson equations \citesuppl{s_umerski_closed-form_1997}
\begin{equation}
	    g^{\mathrm{L}}_{n}=(E-H_{n}-t_{n-1}^{\dagger} g_{n-1}^{\mathrm{L}} t_{n-1})^{-1}
	    \label{add_layer_left}
\end{equation}
and
\begin{equation}
	    g^{\mathrm{R}}_{n}=(E-H_{n}-t_{n} g_{n+1}^{\mathrm{R}}  t_{n}^{\dagger})^{-1}.
	    \label{add_layer_right}
\end{equation}
Further iterations of Eqns.~(\ref{add_layer_left}) and (\ref{add_layer_right}) reduce the problem to the knowledge of the Green's functions at the surfaces separating the scattering region from the left and right leads, $g^{\mathrm{LL}}$ and $g^{\mathrm{RL}}$, that we calculated according to the analytic closed form solution described in Ref.~\citesuppl{s_sanvito_general_1999}.  

The time complexity of the Green's function calculation for each lead with respect to the number $N_{\mathrm{lead}}$ of orbitals in the lead unit cell is $O(N_{\mathrm{lead}}^3)$ \citesuppl{s_umerski_closed-form_1997}. On the other hand, as follows from Eqns.~(\ref{add_layer_left}) and (\ref{add_layer_right}), the complexity of the addition of the layers required to reach the cleavage plane is $O(M \times N_{\rm layer}^3)$, where $M$ and $N_{\rm layer}$ are the number of principal layers and the number of orbitals in each layer, respectively. Consequently, the overall complexity of the method is cubic with respect to the width and linear with respect to the length of the system.

In order to perform our conductance scaling analysis we vary the length of the scattering region in the range $L = 1...60$~nm by steps of 8~PLs, which corresponds to $\Delta L \approx 1$~nm. At each step the right lead is moved rightwards whereas the left lead is kept fixed [Fig.~\ref{FigS2}]. 
The values of conductance $G(E)$ are averaged over an ensemble of $N_{\mathrm{ens}}$=9600 disorder realizations for proper statistical sampling. 

\subsection{Complete account of the results of calculations of conductivity and localization length}

\begin{figure}
\includegraphics[width=13.5cm]{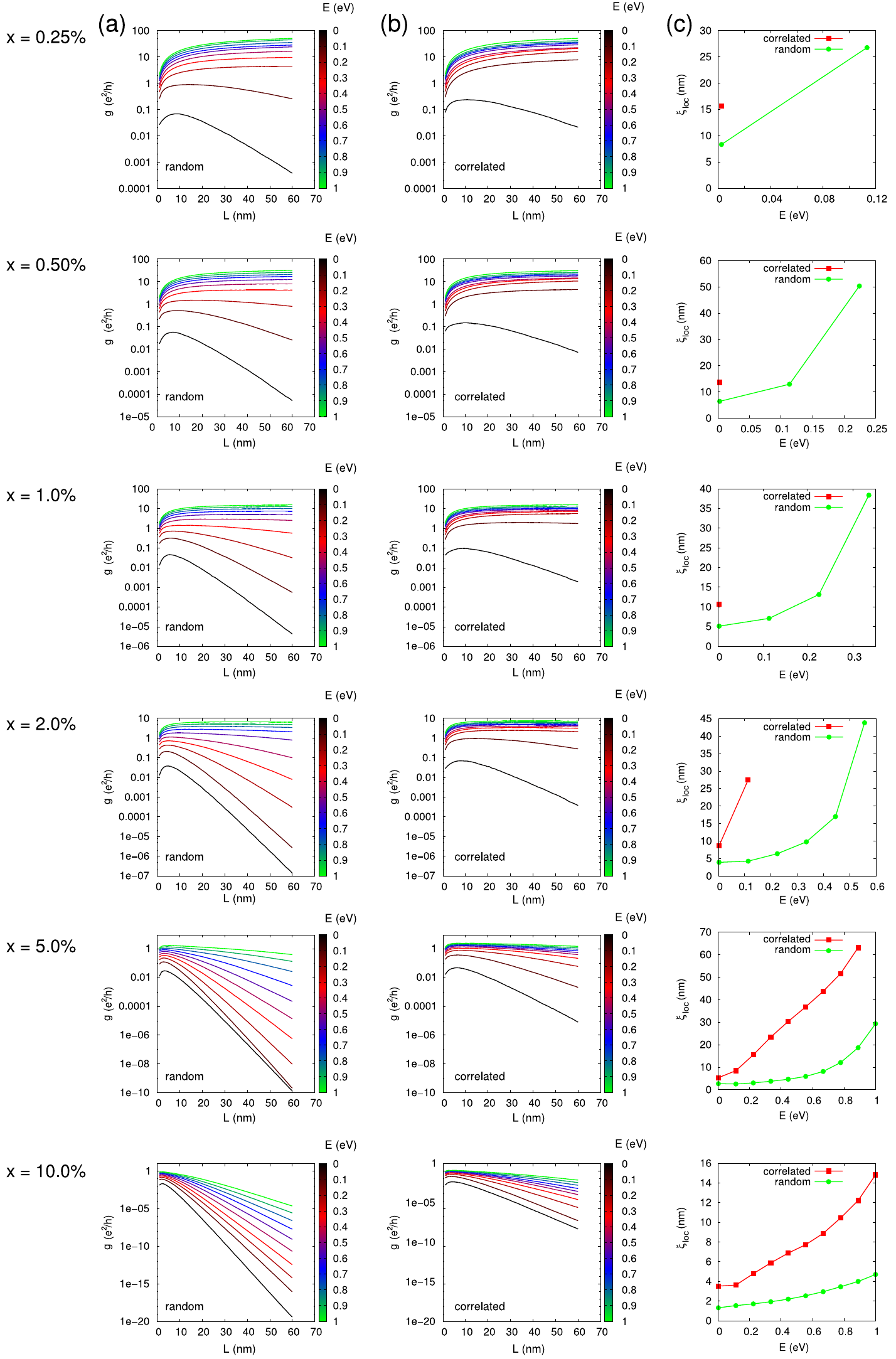}
\caption{(Color online) Scaling analysis of conductivity $g$ and localization length $\xi_{\rm loc}$ for concentrations $x = 0.25\%...10\%$. (a,b) Conductivity $g$ as a function of scattering region length $L$ calculated for graphene with random and correlated adatom distributions, respectively, at charge-carrier energies 0~eV~$< E <$~1~eV. (c) Localization length $\xi_{\rm loc}$ as a function of charge-carrier energy $E$ for random and correlated adatom distributions.}
\label{FigS3}
\end{figure}

Conductivity $g$ for the entire investigated range of concentrations  is presented in Fig.~\ref{FigS3}(a,b). The overall enhancement of conductivity upon the aggregation of adatoms is a common feature at all investigated concentrations. This is particularly visible in the strong localization regime, that is at low energies $E$ and large scattering region lengths $L$. Localization length $\xi_{\rm loc}$ can only be determined for the $g(L)$ curves which exhibit a well-defined negative slope in the large length region. This is the case for $\xi_{\rm loc} < 60$~nm, which is the maximum scattering region length considered in our study. This does not imply that the system does not undergo localization, but rather that a longer scattering region is needed in order to estimate $\xi_{\rm loc}$ correctly. 
For this reason many values of localization length $\xi_{\rm loc}$
at $x<5\%$, especially in the case of correlated impurities, are missing in Fig.~\ref{FigS3}(c).  

%Full_transport

\subsection{Effect of the doping of leads in the Landauer-B\"{u}ttiker calculations}

\begin{figure}[h]
\includegraphics[width=14.5cm]{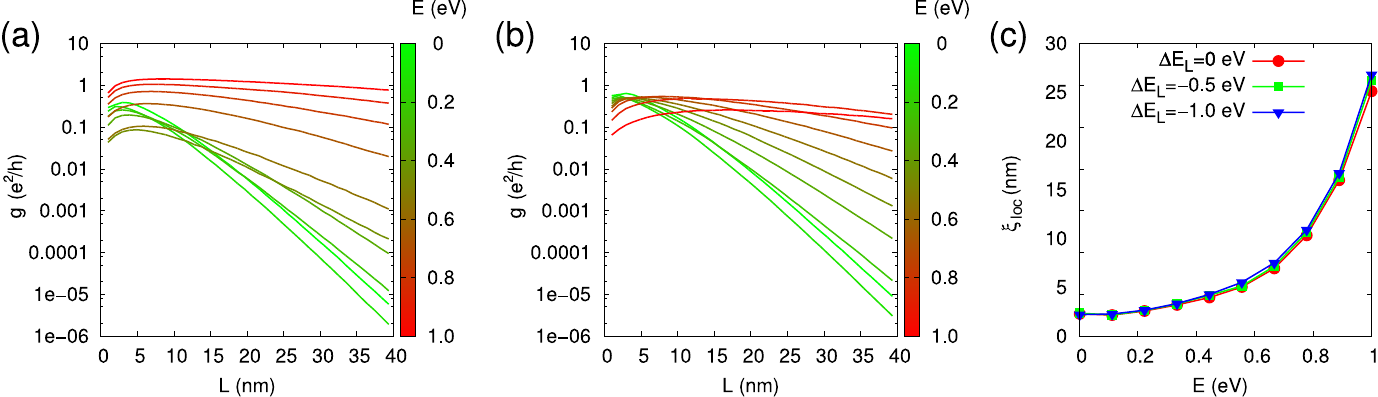}
\caption{(Color online) (a,b) Conductivity $g$ as a function of scattering region length $L$ for graphene with randomly distributed adatoms at $x=5\%$. The charge neutrality point of the leads has been shifted by (a) $\Delta E_{\mathrm{L}}=-0.5$~eV and  (b) $\Delta E_{\mathrm{L}}=-1.0$~eV, respectively. (c) Localization length $\xi_{\mathrm{loc}}$ as a function of charge-carrier energy $E$ for the two investigated lead doping levels compared to undoped leads $\Delta E_{\mathrm{L}}=0$~eV.}
\label{FigS4}
\end{figure}

Figure~\ref{FigS4}(a,b) shows the conductivity curves $g(L)$ for $x=5\%$ concentration of randomly distributed adatoms obtained by shifting the charge neutrality point of leads by $\Delta E_{\mathrm{L}}=-0.5$~eV and $\Delta E_{\mathrm{L}}=-1.0$~eV, respectively. 
The effect of the doping of leads is two-fold. Firstly, the DOS of pristine graphene increases away from the charge neutrality point. Hence, upon doping the number of transport channels increases, which may result in larger values of conductance $g$. This is particularly important when the scattering region has an enhanced DOS at zero energy, such as graphene with resonant impurities. A comparison of Fig.~\ref{FigS4}(a) and Fig.~\ref{FigS3}(a) for $x=5\%$ shows that conductivity is indeed enhanced at low energies with a crossing of the $g(L)$ curves at $L \approx 10$~nm. 
Secondly, doping results in a mismatch between the Fermi wavelength of the leads and that of the scattering region, which has a detrimental effect on conductance $g$. This effect is expected to be more pronounced at higher doping and high energy, where localization plays a smaller role. Indeed, at higher doping ($\Delta E_{\mathrm{L}}=-1.0$~eV) and high energy ($E=1.0$~eV)  the conductivity is reduced, notably at short distance $L<10$~nm [Fig.~\ref{FigS4}(a,b)]. On the other hand, at large distances a general increase of the conductance is progressively restored since the conductivity becomes predominantly determined by the localization of the wavefunction and the increased number of available states. Finally, as shown in Fig.~\ref{FigS4}(c), localization length $\xi_{\mathrm{ loc}}$ is practically unaffected by the doping of the leads, since it is an intrinsic property of the scattering region. 

\subsection{Electronic transport calculations within the Kubo-Greenwood formalism using the kernel polynomial method}

The diagonal elements of the real part of the frequency dependent conductivity tensor in linear response theory is given by the Kubo-Greenwood formula
\begin{equation}
\mathfrak{Re}\lbrace\sigma_{\alpha\alpha}(\omega)\rbrace=\frac{\pi}{V}\int dE\frac{f(E)-f(E+\hbar\omega)}{\omega}\times\mathrm{Tr}\lbrace \delta(E-{H}){j}_{\alpha}\delta(E-{H}+\hbar\omega){j}_{\alpha}\rbrace.
\end{equation}
Here, $f(E)$ is the Fermi function and ${H}$ the single-particle tight-binding Hamiltonian of the system under consideration, whereas the vectorial component of the current operator ${j}_{\alpha}$ is defined later. In the thermodynamic limit for zero temperature and zero frequency this equation reduces to
\begin{equation}
\mathfrak{Re}\lbrace\sigma_{\alpha\alpha}(0)\rbrace=\frac{\pi\hbar}{V}\mathrm{Tr}\lbrace\delta(E_{\rm F}-{H}){j}_{\alpha}\delta(E_{\rm F}-{H}){j}_{\alpha}\rbrace,
\end{equation}
which defines the DC-conductivity studied in this work. The evaluation of this expression is carried out by employing the kernel polynomial method (KPM), as described in detail in Ref.~\citesuppl{s_weise_kernel_2006}. In this framework, the following matrix element density is defined
\begin{equation}
j(E,E^{\prime})=\frac{1}{V}\sum_{n,m}\langle n\vert{j}_{\alpha}\vert m\rangle\langle m\vert{j}_{\alpha}\vert n\rangle\delta(E-\hbar\omega_{n})\delta(E^{\prime}-\hbar\omega_{m}),
\end{equation}
which is then expanded up to finite order $M$ within the two-dimensional KPM using the Jackson kernel. For the studied supercells containing up to $N_{\rm C}\sim10^{6}$ carbon atoms, we used $M=1280$. 
The real-part of the frequency dependent conductivity tensor is then given by a double integration over the matrix density for arbitrary temperature and Fermi level:
\begin{equation}
\mathfrak{Re}\lbrace\sigma_{\alpha\alpha}(\omega)\rbrace=\frac{\pi}{\omega}\int_{-\infty}^{\infty}dE\int_{-\infty}^{\infty}dE^{\prime}j(E,E^{\prime})\times\left[f(E)-f(E^{\prime})\right]\delta(\hbar\omega-(E^{\prime}-E)).
\end{equation}
As a result, the DC-conductivity can be expressed in terms of the diagonal elements of $j(E,E^{\prime})$:
\begin{equation}
\mathfrak{Re}\lbrace\sigma_{\alpha\alpha}(0)\rbrace=\pi\hbar j(E_{\rm F},E_{\rm F}).
\end{equation}
%\subsection*{The current operator}
The calculation of transport properties requires the definition of a current operator appropriate to the employed single-particle tight-binding model. Thus, we approximate the spatial operator ${r}$ in a diagonal form:
\begin{equation}
{r}^{\gamma}\approx\sum\limits_{i}r_{i}^{\gamma}{c}_{i}^{\dagger}{c}_{i}.
\end{equation}
Consequently, one can derive the following current operator used in this work,
\begin{equation}
{j}_{\gamma}=-i\frac{e_0}{\hbar}\sum_{ij}t_{ij}(r_{j}^{\gamma}-r_{i}^{\gamma}){c}_{i}^{\dagger}{c}_{j},
\end{equation}
where indices $i$ and $j$ run over atomic positions and $\gamma$ denotes the vectorial component. For details of a similar derivation see e.g. Ref.~\citesuppl{s_Tomczak09}.

\subsection{Results for the calculations of conductivity using the kernel polynomial method}

\begin{figure}
 \includegraphics[width=9.5cm]{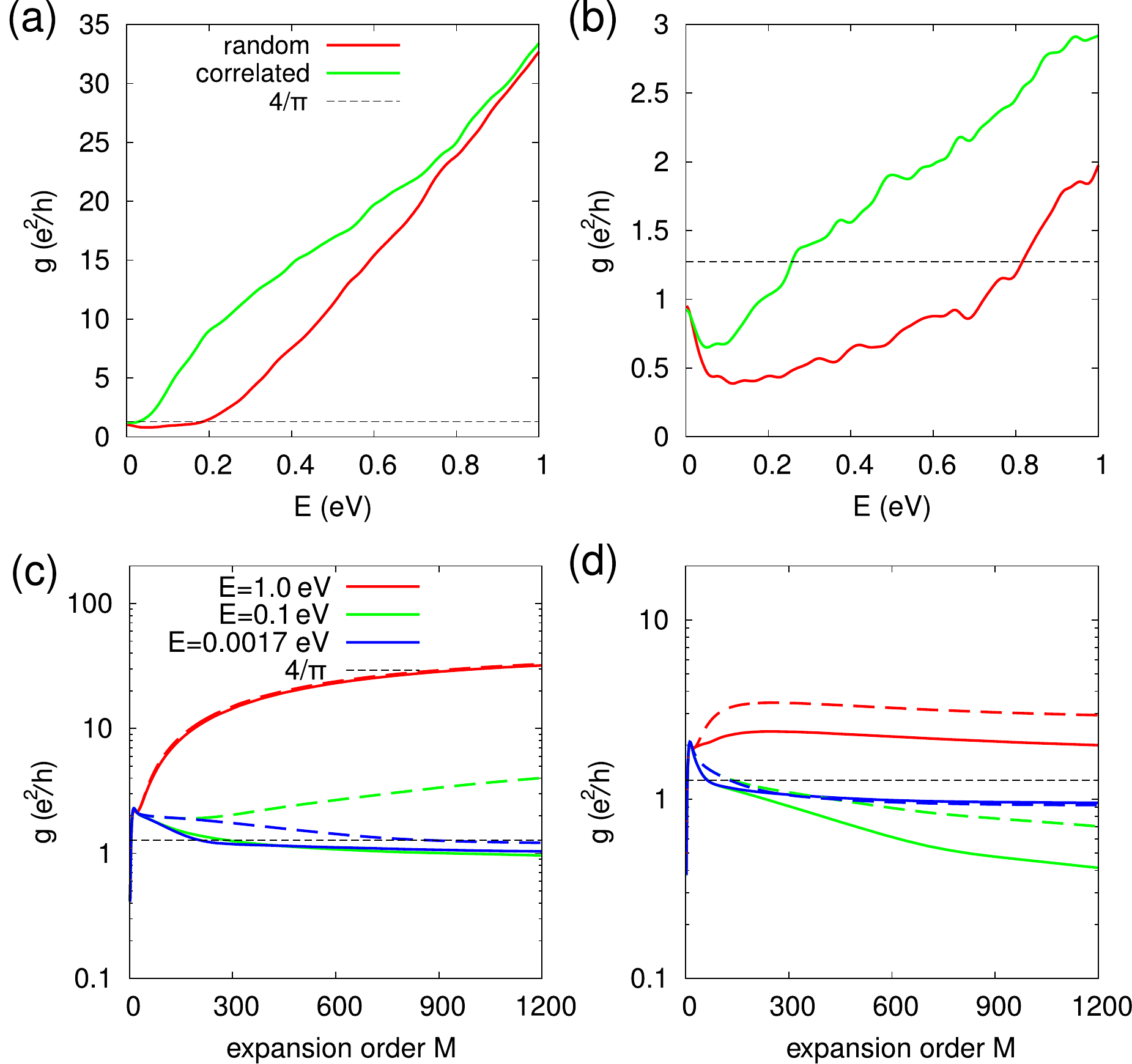}
 \caption{(Color online) (a,b) Conductivity $g$ as a function of charge-carrier energy $E$ calculated using the kernel polynomial method (KPM) for the random and correlated impurity distributions at $x=0.5\%$ and $x=5.0\%$ adatom concentrations, respectively. (c,d) Conductivity $g$ as a function of expansion order $M$ for the random and correlated impurity distributions at $x=0.5\%$ and $x=5.0\%$ adatom concentrations, respectively. The dashed lines indicate the minimum conductivity of graphene.}
 \label{FigS5}
\end{figure}

Figure~\ref{FigS5}(a,b) shows conductivity $g$ as a function of charge-carrier energy $E$ calculated using the KPM for random and correlated impurity distributions at two different adatom concentrations $x=0.5\%$ and $x=5.0\%$.
A scaling analysis of the conductivity for three charge carrier energies ($E = 1.7 \times 10^{-3}$~eV, $E=0.1$~eV and $E=1.0$~eV) based on the expansion order $M$ is shown in Fig.~\ref{FigS5}(c,d) for the same values of adatom concentration $x$. The order $M$ relates to a timescale $\tau(M)$ \citesuppl{s_roche_quantum_1999}. For $E=1.0$~eV and $x=5\%$ the conductivity rises with increasing expansion order $M$, exhibits a maximum around $M\sim 200$ and then decreases slowly again for both correlated (dashed line) and random (solid line) impurity distributions [Fig.~\ref{FigS5}(d)]. We interpret these maxima as the semiclassical values of conductivity $g_{\mathrm{sc}}$ corresponding to the diffusive regime, followed by quantum corrections which result in the decrease of the conductivity. For a concentration of $x=0.5\%$ and $E=1.0$ eV no conductivity maximum is observed, which indicates that the results still correspond to the pre-diffusive/ballistic regime for the largest expansion order investigated. A shift of $g_{\mathrm{sc}}$ to larger expansion orders for decreasing $x$ is expected, because fewer scattering centers are present. The situation is, in general, different for small energies close to the Dirac point (here, $E = 1.7 \times 10^{-3}$~eV), in both cases of correlated and random adatom distributions as well as concentrations of $x=0.5\%$ and $x=5\%$. A distinct maximum for $M>100$ is completely absent, and only a reduction of the conductivity is observed for increasing $M$. Thus, we relate this characteristic to the quantum regime with localization effects resulting in a reduced localization length. If the expansion order is sufficiently large, the curves for $E = 1.7 \times 10^{-3}$~eV reach values well below $g = 4/\pi (e^2/h)$, and one may expect the conductivity to converge asymptotically to zero in the limit of $M\rightarrow\infty$ if the modes are completely localized.

%This is to make sure the bibliography won't be broken by float objects
\FloatBarrier

%Here we put the supp mat bilbiography
\bibliographystylesuppl{apsrev4-1}
\bibliographysuppl{suppl}


\begin{thebibliography}{41}%
\makeatletter
\providecommand \@ifxundefined [1]{%
 \@ifx{#1\undefined}
}%
\providecommand \@ifnum [1]{%
 \ifnum #1\expandafter \@firstoftwo
 \else \expandafter \@secondoftwo
 \fi
}%
\providecommand \@ifx [1]{%
 \ifx #1\expandafter \@firstoftwo
 \else \expandafter \@secondoftwo
 \fi
}%
\providecommand \natexlab [1]{#1}%
\providecommand \enquote  [1]{``#1''}%
\providecommand \bibnamefont  [1]{#1}%
\providecommand \bibfnamefont [1]{#1}%
\providecommand \citenamefont [1]{#1}%
\providecommand \href@noop [0]{\@secondoftwo}%
\providecommand \href [0]{\begingroup \@sanitize@url \@href}%
\providecommand \@href[1]{\@@startlink{#1}\@@href}%
\providecommand \@@href[1]{\endgroup#1\@@endlink}%
\providecommand \@sanitize@url [0]{\catcode `\\12\catcode `\$12\catcode
  `\&12\catcode `\#12\catcode `\^12\catcode `\_12\catcode `\%12\relax}%
\providecommand \@@startlink[1]{}%
\providecommand \@@endlink[0]{}%
\providecommand \url  [0]{\begingroup\@sanitize@url \@url }%
\providecommand \@url [1]{\endgroup\@href {#1}{\urlprefix }}%
\providecommand \urlprefix  [0]{URL }%
\providecommand \Eprint [0]{\href }%
\providecommand \doibase [0]{http://dx.doi.org/}%
\providecommand \selectlanguage [0]{\@gobble}%
\providecommand \bibinfo  [0]{\@secondoftwo}%
\providecommand \bibfield  [0]{\@secondoftwo}%
\providecommand \translation [1]{[#1]}%
\providecommand \BibitemOpen [0]{}%
\providecommand \bibitemStop [0]{}%
\providecommand \bibitemNoStop [0]{.\EOS\space}%
\providecommand \EOS [0]{\spacefactor3000\relax}%
\providecommand \BibitemShut  [1]{\csname bibitem#1\endcsname}%
\let\auto@bib@innerbib\@empty
%</preamble>
\bibitem [{\citenamefont {Geim}\ and\ \citenamefont
  {Novoselov}(2007)}]{geim_rise_2007}%
  \BibitemOpen
  \bibfield  {author} {\bibinfo {author} {\bibfnamefont {A.~K.}\ \bibnamefont
  {Geim}}\ and\ \bibinfo {author} {\bibfnamefont {K.~S.}\ \bibnamefont
  {Novoselov}},\ }\href@noop {} {\bibfield  {journal} {\bibinfo  {journal}
  {Nature Mater.}\ }\textbf {\bibinfo {volume} {6}},\ \bibinfo {pages} {183}
  (\bibinfo {year} {2007})}\BibitemShut {NoStop}%
\bibitem [{\citenamefont {Castro~Neto}\ \emph {et~al.}(2009)\citenamefont
  {Castro~Neto}, \citenamefont {Guinea}, \citenamefont {Peres}, \citenamefont
  {Novoselov},\ and\ \citenamefont {Geim}}]{castro_neto_electronic_2009}%
  \BibitemOpen
  \bibfield  {author} {\bibinfo {author} {\bibfnamefont {A.~H.}\ \bibnamefont
  {Castro~Neto}}, \bibinfo {author} {\bibfnamefont {F.}~\bibnamefont {Guinea}},
  \bibinfo {author} {\bibfnamefont {N.~M.~R.}\ \bibnamefont {Peres}}, \bibinfo
  {author} {\bibfnamefont {K.~S.}\ \bibnamefont {Novoselov}}, \ and\ \bibinfo
  {author} {\bibfnamefont {A.~K.}\ \bibnamefont {Geim}},\ }\href@noop {}
  {\bibfield  {journal} {\bibinfo  {journal} {Rev. Mod. Phys.}\ }\textbf
  {\bibinfo {volume} {81}},\ \bibinfo {pages} {109} (\bibinfo {year}
  {2009})}\BibitemShut {NoStop}%
\bibitem [{\citenamefont {Das~Sarma}\ \emph {et~al.}(2011)\citenamefont
  {Das~Sarma}, \citenamefont {Adam}, \citenamefont {Hwang},\ and\ \citenamefont
  {Rossi}}]{transport_review_graphene_2011}%
  \BibitemOpen
  \bibfield  {author} {\bibinfo {author} {\bibfnamefont {S.}~\bibnamefont
  {Das~Sarma}}, \bibinfo {author} {\bibfnamefont {S.}~\bibnamefont {Adam}},
  \bibinfo {author} {\bibfnamefont {E.~H.}\ \bibnamefont {Hwang}}, \ and\
  \bibinfo {author} {\bibfnamefont {E.}~\bibnamefont {Rossi}},\ }\href@noop {}
  {\bibfield  {journal} {\bibinfo  {journal} {Rev. Mod. Phys.}\ }\textbf
  {\bibinfo {volume} {83}},\ \bibinfo {pages} {407} (\bibinfo {year}
  {2011})}\BibitemShut {NoStop}%
\bibitem [{\citenamefont {Tworzydło}\ \emph {et~al.}(2006)\citenamefont
  {Tworzydło}, \citenamefont {Trauzettel}, \citenamefont {Titov},
  \citenamefont {Rycerz},\ and\ \citenamefont
  {Beenakker}}]{tworzydlo_sub-poissonian_2006}%
  \BibitemOpen
  \bibfield  {author} {\bibinfo {author} {\bibfnamefont {J.}~\bibnamefont
  {Tworzydło}}, \bibinfo {author} {\bibfnamefont {B.}~\bibnamefont
  {Trauzettel}}, \bibinfo {author} {\bibfnamefont {M.}~\bibnamefont {Titov}},
  \bibinfo {author} {\bibfnamefont {A.}~\bibnamefont {Rycerz}}, \ and\ \bibinfo
  {author} {\bibfnamefont {C.~W.~J.}\ \bibnamefont {Beenakker}},\ }\href@noop
  {} {\bibfield  {journal} {\bibinfo  {journal} {Phys. Rev. Lett.}\ }\textbf
  {\bibinfo {volume} {96}},\ \bibinfo {pages} {246802} (\bibinfo {year}
  {2006})}\BibitemShut {NoStop}%
\bibitem [{\citenamefont {Katsnelson}\ \emph {et~al.}(2006)\citenamefont
  {Katsnelson}, \citenamefont {Novoselov},\ and\ \citenamefont
  {Geim}}]{katsnelson_chiral_2006}%
  \BibitemOpen
  \bibfield  {author} {\bibinfo {author} {\bibfnamefont {M.~I.}\ \bibnamefont
  {Katsnelson}}, \bibinfo {author} {\bibfnamefont {K.~S.}\ \bibnamefont
  {Novoselov}}, \ and\ \bibinfo {author} {\bibfnamefont {A.~K.}\ \bibnamefont
  {Geim}},\ }\href@noop {} {\bibfield  {journal} {\bibinfo  {journal} {Nature
  Phys.}\ }\textbf {\bibinfo {volume} {2}},\ \bibinfo {pages} {620} (\bibinfo
  {year} {2006})}\BibitemShut {NoStop}%
\bibitem [{\citenamefont {Zhang}\ \emph {et~al.}(2005)\citenamefont {Zhang},
  \citenamefont {Tan}, \citenamefont {Stormer},\ and\ \citenamefont
  {Kim}}]{zhang_experimental_2005}%
  \BibitemOpen
  \bibfield  {author} {\bibinfo {author} {\bibfnamefont {Y.}~\bibnamefont
  {Zhang}}, \bibinfo {author} {\bibfnamefont {Y.-W.}\ \bibnamefont {Tan}},
  \bibinfo {author} {\bibfnamefont {H.~L.}\ \bibnamefont {Stormer}}, \ and\
  \bibinfo {author} {\bibfnamefont {P.}~\bibnamefont {Kim}},\ }\href@noop {}
  {\bibfield  {journal} {\bibinfo  {journal} {Nature (London)}\ }\textbf
  {\bibinfo {volume} {438}},\ \bibinfo {pages} {201} (\bibinfo {year}
  {2005})}\BibitemShut {NoStop}%
\bibitem [{\citenamefont {Novoselov}\ \emph {et~al.}(2005)\citenamefont
  {Novoselov}, \citenamefont {Geim}, \citenamefont {Morozov}, \citenamefont
  {Jiang}, \citenamefont {Katsnelson}, \citenamefont {Grigorieva},
  \citenamefont {Dubonos},\ and\ \citenamefont
  {Firsov}}]{novoselov_two-dimensional_2005}%
  \BibitemOpen
  \bibfield  {author} {\bibinfo {author} {\bibfnamefont {K.~S.}\ \bibnamefont
  {Novoselov}}, \bibinfo {author} {\bibfnamefont {A.~K.}\ \bibnamefont {Geim}},
  \bibinfo {author} {\bibfnamefont {S.~V.}\ \bibnamefont {Morozov}}, \bibinfo
  {author} {\bibfnamefont {D.}~\bibnamefont {Jiang}}, \bibinfo {author}
  {\bibfnamefont {M.~I.}\ \bibnamefont {Katsnelson}}, \bibinfo {author}
  {\bibfnamefont {I.~V.}\ \bibnamefont {Grigorieva}}, \bibinfo {author}
  {\bibfnamefont {S.~V.}\ \bibnamefont {Dubonos}}, \ and\ \bibinfo {author}
  {\bibfnamefont {A.~A.}\ \bibnamefont {Firsov}},\ }\href@noop {} {\bibfield
  {journal} {\bibinfo  {journal} {Nature (London)}\ }\textbf {\bibinfo {volume}
  {438}},\ \bibinfo {pages} {197} (\bibinfo {year} {2005})}\BibitemShut
  {NoStop}%
\bibitem [{\citenamefont {Wehling}\ \emph {et~al.}(2010)\citenamefont
  {Wehling}, \citenamefont {Yuan}, \citenamefont {Lichtenstein}, \citenamefont
  {Geim},\ and\ \citenamefont {Katsnelson}}]{wehling_resonant_2010}%
  \BibitemOpen
  \bibfield  {author} {\bibinfo {author} {\bibfnamefont {T.~O.}\ \bibnamefont
  {Wehling}}, \bibinfo {author} {\bibfnamefont {S.}~\bibnamefont {Yuan}},
  \bibinfo {author} {\bibfnamefont {A.~I.}\ \bibnamefont {Lichtenstein}},
  \bibinfo {author} {\bibfnamefont {A.~K.}\ \bibnamefont {Geim}}, \ and\
  \bibinfo {author} {\bibfnamefont {M.~I.}\ \bibnamefont {Katsnelson}},\
  }\href@noop {} {\bibfield  {journal} {\bibinfo  {journal} {Phys. Rev. Lett.}\
  }\textbf {\bibinfo {volume} {105}},\ \bibinfo {pages} {056802} (\bibinfo
  {year} {2010})}\BibitemShut {NoStop}%
\bibitem [{\citenamefont {Gargiulo}\ and\ \citenamefont
  {Yazyev}(2014)}]{gargiulo_topological_2013}%
  \BibitemOpen
  \bibfield  {author} {\bibinfo {author} {\bibfnamefont {F.}~\bibnamefont
  {Gargiulo}}\ and\ \bibinfo {author} {\bibfnamefont {O.~V.}\ \bibnamefont
  {Yazyev}},\ }\href@noop {} {\bibfield  {journal} {\bibinfo  {journal} {Nano
  Lett.}\ }\textbf {\bibinfo {volume} {14}},\ \bibinfo {pages} {250} (\bibinfo
  {year} {2014})}\BibitemShut {NoStop}%
\bibitem [{\citenamefont {Ni}\ \emph {et~al.}(2010)\citenamefont {Ni},
  \citenamefont {Ponomarenko}, \citenamefont {Nair}, \citenamefont {Yang},
  \citenamefont {Anissimova}, \citenamefont {Grigorieva}, \citenamefont
  {Schedin}, \citenamefont {Blake}, \citenamefont {Shen}, \citenamefont {Hill},
  \citenamefont {Novoselov},\ and\ \citenamefont {Geim}}]{ni_resonant_2010}%
  \BibitemOpen
  \bibfield  {author} {\bibinfo {author} {\bibfnamefont {Z.~H.}\ \bibnamefont
  {Ni}}, \bibinfo {author} {\bibfnamefont {L.~A.}\ \bibnamefont {Ponomarenko}},
  \bibinfo {author} {\bibfnamefont {R.~R.}\ \bibnamefont {Nair}}, \bibinfo
  {author} {\bibfnamefont {R.}~\bibnamefont {Yang}}, \bibinfo {author}
  {\bibfnamefont {S.}~\bibnamefont {Anissimova}}, \bibinfo {author}
  {\bibfnamefont {I.~V.}\ \bibnamefont {Grigorieva}}, \bibinfo {author}
  {\bibfnamefont {F.}~\bibnamefont {Schedin}}, \bibinfo {author} {\bibfnamefont
  {P.}~\bibnamefont {Blake}}, \bibinfo {author} {\bibfnamefont {Z.~X.}\
  \bibnamefont {Shen}}, \bibinfo {author} {\bibfnamefont {E.~H.}\ \bibnamefont
  {Hill}}, \bibinfo {author} {\bibfnamefont {K.~S.}\ \bibnamefont {Novoselov}},
  \ and\ \bibinfo {author} {\bibfnamefont {A.~K.}\ \bibnamefont {Geim}},\
  }\href@noop {} {\bibfield  {journal} {\bibinfo  {journal} {Nano Lett.}\
  }\textbf {\bibinfo {volume} {10}},\ \bibinfo {pages} {3868} (\bibinfo {year}
  {2010})}\BibitemShut {NoStop}%
\bibitem [{\citenamefont {Wehling}\ \emph {et~al.}(2007)\citenamefont
  {Wehling}, \citenamefont {Balatsky}, \citenamefont {Katsnelson},
  \citenamefont {Lichtenstein}, \citenamefont {Scharnberg},\ and\ \citenamefont
  {Wiesendanger}}]{wehling_local_2007}%
  \BibitemOpen
  \bibfield  {author} {\bibinfo {author} {\bibfnamefont {T.~O.}\ \bibnamefont
  {Wehling}}, \bibinfo {author} {\bibfnamefont {A.~V.}\ \bibnamefont
  {Balatsky}}, \bibinfo {author} {\bibfnamefont {M.~I.}\ \bibnamefont
  {Katsnelson}}, \bibinfo {author} {\bibfnamefont {A.~I.}\ \bibnamefont
  {Lichtenstein}}, \bibinfo {author} {\bibfnamefont {K.}~\bibnamefont
  {Scharnberg}}, \ and\ \bibinfo {author} {\bibfnamefont {R.}~\bibnamefont
  {Wiesendanger}},\ }\href@noop {} {\bibfield  {journal} {\bibinfo  {journal}
  {Phys. Rev. B}\ }\textbf {\bibinfo {volume} {75}},\ \bibinfo {pages} {125425}
  (\bibinfo {year} {2007})}\BibitemShut {NoStop}%
\bibitem [{\citenamefont {Kramer}\ and\ \citenamefont
  {{MacKinnon}}(1993)}]{kramer_localization:_1993}%
  \BibitemOpen
  \bibfield  {author} {\bibinfo {author} {\bibfnamefont {B.}~\bibnamefont
  {Kramer}}\ and\ \bibinfo {author} {\bibfnamefont {A.}~\bibnamefont
  {{MacKinnon}}},\ }\href@noop {} {\bibfield  {journal} {\bibinfo  {journal}
  {Rep. Prog. Phys.}\ }\textbf {\bibinfo {volume} {56}},\ \bibinfo {pages}
  {1469} (\bibinfo {year} {1993})}\BibitemShut {NoStop}%
\bibitem [{\citenamefont {Evers}\ and\ \citenamefont
  {Mirlin}(2008)}]{evers_anderson_2008}%
  \BibitemOpen
  \bibfield  {author} {\bibinfo {author} {\bibfnamefont {F.}~\bibnamefont
  {Evers}}\ and\ \bibinfo {author} {\bibfnamefont {A.~D.}\ \bibnamefont
  {Mirlin}},\ }\href@noop {} {\bibfield  {journal} {\bibinfo  {journal} {Rev.
  Mod. Phys.}\ }\textbf {\bibinfo {volume} {80}},\ \bibinfo {pages} {1355}
  (\bibinfo {year} {2008})}\BibitemShut {NoStop}%
\bibitem [{\citenamefont {Bostwick}\ \emph {et~al.}(2009)\citenamefont
  {Bostwick}, \citenamefont {McChesney}, \citenamefont {Emtsev}, \citenamefont
  {Seyller}, \citenamefont {Horn}, \citenamefont {Kevan},\ and\ \citenamefont
  {Rotenberg}}]{bostwick_quasiparticle_2009}%
  \BibitemOpen
  \bibfield  {author} {\bibinfo {author} {\bibfnamefont {A.}~\bibnamefont
  {Bostwick}}, \bibinfo {author} {\bibfnamefont {J.~L.}\ \bibnamefont
  {McChesney}}, \bibinfo {author} {\bibfnamefont {K.~V.}\ \bibnamefont
  {Emtsev}}, \bibinfo {author} {\bibfnamefont {T.}~\bibnamefont {Seyller}},
  \bibinfo {author} {\bibfnamefont {K.}~\bibnamefont {Horn}}, \bibinfo {author}
  {\bibfnamefont {S.~D.}\ \bibnamefont {Kevan}}, \ and\ \bibinfo {author}
  {\bibfnamefont {E.}~\bibnamefont {Rotenberg}},\ }\href@noop {} {\bibfield
  {journal} {\bibinfo  {journal} {Phys. Rev. Lett.}\ }\textbf {\bibinfo
  {volume} {103}},\ \bibinfo {pages} {056404} (\bibinfo {year}
  {2009})}\BibitemShut {NoStop}%
\bibitem [{\citenamefont {Song}\ \emph {et~al.}(2011)\citenamefont {Song},
  \citenamefont {Song},\ and\ \citenamefont {Feng}}]{song_effects_2011}%
  \BibitemOpen
  \bibfield  {author} {\bibinfo {author} {\bibfnamefont {Y.}~\bibnamefont
  {Song}}, \bibinfo {author} {\bibfnamefont {H.}~\bibnamefont {Song}}, \ and\
  \bibinfo {author} {\bibfnamefont {S.}~\bibnamefont {Feng}},\ }\href@noop {}
  {\bibfield  {journal} {\bibinfo  {journal} {J. Phys.: Condens. Matter}\
  }\textbf {\bibinfo {volume} {23}},\ \bibinfo {pages} {205501} (\bibinfo
  {year} {2011})}\BibitemShut {NoStop}%
\bibitem [{\citenamefont {Jayasingha}\ \emph {et~al.}(2013)\citenamefont
  {Jayasingha}, \citenamefont {Sherehiy}, \citenamefont {Wu},\ and\
  \citenamefont {Sumanasekera}}]{jayasingha_situ_2013}%
  \BibitemOpen
  \bibfield  {author} {\bibinfo {author} {\bibfnamefont {R.}~\bibnamefont
  {Jayasingha}}, \bibinfo {author} {\bibfnamefont {A.}~\bibnamefont
  {Sherehiy}}, \bibinfo {author} {\bibfnamefont {S.-Y.}\ \bibnamefont {Wu}}, \
  and\ \bibinfo {author} {\bibfnamefont {G.~U.}\ \bibnamefont {Sumanasekera}},\
  }\href@noop {} {\bibfield  {journal} {\bibinfo  {journal} {Nano Lett.}\
  }\textbf {\bibinfo {volume} {13}},\ \bibinfo {pages} {5098} (\bibinfo {year}
  {2013})}\BibitemShut {NoStop}%
\bibitem [{\citenamefont {Adam}\ \emph {et~al.}(2008)\citenamefont {Adam},
  \citenamefont {Cho}, \citenamefont {Fuhrer},\ and\ \citenamefont
  {Das~Sarma}}]{adam_density_2008}%
  \BibitemOpen
  \bibfield  {author} {\bibinfo {author} {\bibfnamefont {S.}~\bibnamefont
  {Adam}}, \bibinfo {author} {\bibfnamefont {S.}~\bibnamefont {Cho}}, \bibinfo
  {author} {\bibfnamefont {M.~S.}\ \bibnamefont {Fuhrer}}, \ and\ \bibinfo
  {author} {\bibfnamefont {S.}~\bibnamefont {Das~Sarma}},\ }\href@noop {}
  {\bibfield  {journal} {\bibinfo  {journal} {Phys. Rev. Lett.}\ }\textbf
  {\bibinfo {volume} {101}},\ \bibinfo {pages} {046404} (\bibinfo {year}
  {2008})}\BibitemShut {NoStop}%
\bibitem [{\citenamefont {Shon}\ and\ \citenamefont
  {Ando}(1998)}]{shon_quantum_1998}%
  \BibitemOpen
  \bibfield  {author} {\bibinfo {author} {\bibfnamefont {N.}~\bibnamefont
  {Shon}}\ and\ \bibinfo {author} {\bibfnamefont {T.}~\bibnamefont {Ando}},\
  }\href@noop {} {\bibfield  {journal} {\bibinfo  {journal} {J. Phys. Soc.
  Jpn.}\ }\textbf {\bibinfo {volume} {67}},\ \bibinfo {pages} {2421} (\bibinfo
  {year} {1998})}\BibitemShut {NoStop}%
\bibitem [{\citenamefont {Wehling}\ \emph {et~al.}(2009)\citenamefont
  {Wehling}, \citenamefont {Katsnelson},\ and\ \citenamefont
  {Lichtenstein}}]{wehling_impurities_2009}%
  \BibitemOpen
  \bibfield  {author} {\bibinfo {author} {\bibfnamefont {T.~O.}\ \bibnamefont
  {Wehling}}, \bibinfo {author} {\bibfnamefont {M.~I.}\ \bibnamefont
  {Katsnelson}}, \ and\ \bibinfo {author} {\bibfnamefont {A.~I.}\ \bibnamefont
  {Lichtenstein}},\ }\href@noop {} {\bibfield  {journal} {\bibinfo  {journal}
  {Phys. Rev. B}\ }\textbf {\bibinfo {volume} {80}},\ \bibinfo {pages} {085428}
  (\bibinfo {year} {2009})}\BibitemShut {NoStop}%
\bibitem [{\citenamefont {Ferreira}\ \emph {et~al.}(2011)\citenamefont
  {Ferreira}, \citenamefont {Viana-Gomes}, \citenamefont {Nilsson},
  \citenamefont {Mucciolo}, \citenamefont {Peres},\ and\ \citenamefont
  {Castro~Neto}}]{ferreira_unified_2011}%
  \BibitemOpen
  \bibfield  {author} {\bibinfo {author} {\bibfnamefont {A.}~\bibnamefont
  {Ferreira}}, \bibinfo {author} {\bibfnamefont {J.}~\bibnamefont
  {Viana-Gomes}}, \bibinfo {author} {\bibfnamefont {J.}~\bibnamefont
  {Nilsson}}, \bibinfo {author} {\bibfnamefont {E.~R.}\ \bibnamefont
  {Mucciolo}}, \bibinfo {author} {\bibfnamefont {N.~M.~R.}\ \bibnamefont
  {Peres}}, \ and\ \bibinfo {author} {\bibfnamefont {A.~H.}\ \bibnamefont
  {Castro~Neto}},\ }\href@noop {} {\bibfield  {journal} {\bibinfo  {journal}
  {Phys. Rev. B}\ }\textbf {\bibinfo {volume} {83}},\ \bibinfo {pages} {165402}
  (\bibinfo {year} {2011})}\BibitemShut {NoStop}%
\bibitem [{\citenamefont {Cresti}\ \emph {et~al.}(2013)\citenamefont {Cresti},
  \citenamefont {Ortmann}, \citenamefont {Louvet}, \citenamefont {Van~Tuan},\
  and\ \citenamefont {Roche}}]{Cresti2013}%
  \BibitemOpen
  \bibfield  {author} {\bibinfo {author} {\bibfnamefont {A.}~\bibnamefont
  {Cresti}}, \bibinfo {author} {\bibfnamefont {F.}~\bibnamefont {Ortmann}},
  \bibinfo {author} {\bibfnamefont {T.}~\bibnamefont {Louvet}}, \bibinfo
  {author} {\bibfnamefont {D.}~\bibnamefont {Van~Tuan}}, \ and\ \bibinfo
  {author} {\bibfnamefont {S.}~\bibnamefont {Roche}},\ }\href@noop {}
  {\bibfield  {journal} {\bibinfo  {journal} {Phys. Rev. Lett.}\ }\textbf
  {\bibinfo {volume} {110}},\ \bibinfo {pages} {196601} (\bibinfo {year}
  {2013})}\BibitemShut {NoStop}%
\bibitem [{\citenamefont {Palacios}\ \emph {et~al.}(2008)\citenamefont
  {Palacios}, \citenamefont {Fernández-Rossier},\ and\ \citenamefont
  {Brey}}]{palacios_vacancy-induced_2008}%
  \BibitemOpen
  \bibfield  {author} {\bibinfo {author} {\bibfnamefont {J.~J.}\ \bibnamefont
  {Palacios}}, \bibinfo {author} {\bibfnamefont {J.}~\bibnamefont
  {Fernández-Rossier}}, \ and\ \bibinfo {author} {\bibfnamefont
  {L.}~\bibnamefont {Brey}},\ }\href@noop {} {\bibfield  {journal} {\bibinfo
  {journal} {Phys. Rev. B}\ }\textbf {\bibinfo {volume} {77}},\ \bibinfo
  {pages} {195428} (\bibinfo {year} {2008})}\BibitemShut {NoStop}%
\bibitem [{\citenamefont {Leconte}\ \emph {et~al.}(2011)\citenamefont
  {Leconte}, \citenamefont {Soriano}, \citenamefont {Roche}, \citenamefont
  {Ordejon}, \citenamefont {Charlier},\ and\ \citenamefont
  {Palacios}}]{leconte_magnetism-dependent_2011}%
  \BibitemOpen
  \bibfield  {author} {\bibinfo {author} {\bibfnamefont {N.}~\bibnamefont
  {Leconte}}, \bibinfo {author} {\bibfnamefont {D.}~\bibnamefont {Soriano}},
  \bibinfo {author} {\bibfnamefont {S.}~\bibnamefont {Roche}}, \bibinfo
  {author} {\bibfnamefont {P.}~\bibnamefont {Ordejon}}, \bibinfo {author}
  {\bibfnamefont {J.-C.}\ \bibnamefont {Charlier}}, \ and\ \bibinfo {author}
  {\bibfnamefont {J.~J.}\ \bibnamefont {Palacios}},\ }\href@noop {} {\bibfield
  {journal} {\bibinfo  {journal} {ACS Nano}\ }\textbf {\bibinfo {volume} {5}},\
  \bibinfo {pages} {3987} (\bibinfo {year} {2011})}\BibitemShut {NoStop}%
\bibitem [{\citenamefont {Trambly~de Laissardière}\ and\ \citenamefont
  {Mayou}(2013)}]{trambly_de_laissardiere_conductivity_2013}%
  \BibitemOpen
  \bibfield  {author} {\bibinfo {author} {\bibfnamefont {G.}~\bibnamefont
  {Trambly~de Laissardière}}\ and\ \bibinfo {author} {\bibfnamefont
  {D.}~\bibnamefont {Mayou}},\ }\href@noop {} {\bibfield  {journal} {\bibinfo
  {journal} {Phys. Rev. Lett.}\ }\textbf {\bibinfo {volume} {111}},\ \bibinfo
  {pages} {146601} (\bibinfo {year} {2013})}\BibitemShut {NoStop}%
\bibitem [{\citenamefont {Boukhvalov}\ \emph {et~al.}(2008)\citenamefont
  {Boukhvalov}, \citenamefont {Katsnelson},\ and\ \citenamefont
  {Lichtenstein}}]{boukhvalov_hydrogen_2008}%
  \BibitemOpen
  \bibfield  {author} {\bibinfo {author} {\bibfnamefont {D.~W.}\ \bibnamefont
  {Boukhvalov}}, \bibinfo {author} {\bibfnamefont {M.~I.}\ \bibnamefont
  {Katsnelson}}, \ and\ \bibinfo {author} {\bibfnamefont {A.~I.}\ \bibnamefont
  {Lichtenstein}},\ }\href@noop {} {\bibfield  {journal} {\bibinfo  {journal}
  {Phys. Rev. B}\ }\textbf {\bibinfo {volume} {77}},\ \bibinfo {pages} {035427}
  (\bibinfo {year} {2008})}\BibitemShut {NoStop}%
\bibitem [{\citenamefont {Lin}\ \emph {et~al.}(2008)\citenamefont {Lin},
  \citenamefont {Ding},\ and\ \citenamefont {Yakobson}}]{lin_hydrogen_2008}%
  \BibitemOpen
  \bibfield  {author} {\bibinfo {author} {\bibfnamefont {Y.}~\bibnamefont
  {Lin}}, \bibinfo {author} {\bibfnamefont {F.}~\bibnamefont {Ding}}, \ and\
  \bibinfo {author} {\bibfnamefont {B.~I.}\ \bibnamefont {Yakobson}},\
  }\href@noop {} {\bibfield  {journal} {\bibinfo  {journal} {Phys. Rev. B}\
  }\textbf {\bibinfo {volume} {78}},\ \bibinfo {pages} {041402} (\bibinfo
  {year} {2008})}\BibitemShut {NoStop}%
\bibitem [{\citenamefont {Yazyev}(2008)}]{Yazyev2008}%
  \BibitemOpen
  \bibfield  {author} {\bibinfo {author} {\bibfnamefont {O.~V.}\ \bibnamefont
  {Yazyev}},\ }\href@noop {} {\bibfield  {journal} {\bibinfo  {journal} {Phys.
  Rev. Lett.}\ }\textbf {\bibinfo {volume} {101}},\ \bibinfo {pages} {037203}
  (\bibinfo {year} {2008})}\BibitemShut {NoStop}%
\bibitem [{\citenamefont {Herrero}\ and\ \citenamefont
  {Ramírez}(2009)}]{herrero_vibrational_2009}%
  \BibitemOpen
  \bibfield  {author} {\bibinfo {author} {\bibfnamefont {C.~P.}\ \bibnamefont
  {Herrero}}\ and\ \bibinfo {author} {\bibfnamefont {R.}~\bibnamefont
  {Ramírez}},\ }\href@noop {} {\bibfield  {journal} {\bibinfo  {journal}
  {Phys. Rev. B}\ }\textbf {\bibinfo {volume} {79}},\ \bibinfo {pages} {115429}
  (\bibinfo {year} {2009})}\BibitemShut {NoStop}%
\bibitem [{\citenamefont {Moaied}\ \emph {et~al.}(2014)\citenamefont {Moaied},
  \citenamefont {Moreno}, \citenamefont {Caturla},\ and\ \citenamefont
  {Palacios}}]{moaied_theoretical_2014}%
  \BibitemOpen
  \bibfield  {author} {\bibinfo {author} {\bibfnamefont {M.}~\bibnamefont
  {Moaied}}, \bibinfo {author} {\bibfnamefont {J.~A.}\ \bibnamefont {Moreno}},
  \bibinfo {author} {\bibfnamefont {M.~J.}\ \bibnamefont {Caturla}}, \ and\
  \bibinfo {author} {\bibfnamefont {J.~J.}\ \bibnamefont {Palacios}},\
  }\href@noop {} {\bibfield  {journal} {\bibinfo  {journal} {arXiv:1405.3165}\
  } (\bibinfo {year} {2014})}\BibitemShut {NoStop}%
\bibitem [{\citenamefont {Yazyev}\ and\ \citenamefont
  {Helm}(2007)}]{Yazyev2007}%
  \BibitemOpen
  \bibfield  {author} {\bibinfo {author} {\bibfnamefont {O.~V.}\ \bibnamefont
  {Yazyev}}\ and\ \bibinfo {author} {\bibfnamefont {L.}~\bibnamefont {Helm}},\
  }\href@noop {} {\bibfield  {journal} {\bibinfo  {journal} {Phys. Rev. B}\
  }\textbf {\bibinfo {volume} {75}},\ \bibinfo {pages} {125408} (\bibinfo
  {year} {2007})}\BibitemShut {NoStop}%
\bibitem [{sup()}]{suppl}%
  \BibitemOpen
  \href@noop {} {}\bibinfo {note} {See Supplemental Material at [URL will be
  inserted by publisher] for detailed description of the methodology and
  additional calculations.}\BibitemShut {Stop}%
\bibitem [{not()}]{note_both_sides}%
  \BibitemOpen
  \href@noop {} {}\bibinfo {note} {In the case of hydrogen adsorption on both
  sides of graphene the fitted parameters $\gamma_1 = -1.461$~eV and $\gamma_2
  = 0.342$~eV. The resulting cluster size distributions, however, are very
  close to those obtained for single-side adsorption \cite{suppl}.}\BibitemShut
  {Stop}%
\bibitem [{\citenamefont {Metropolis}\ \emph {et~al.}(1953)\citenamefont
  {Metropolis}, \citenamefont {Rosenbluth}, \citenamefont {Rosenbluth},
  \citenamefont {Teller},\ and\ \citenamefont
  {Teller}}]{metropolis_equation_1953}%
  \BibitemOpen
  \bibfield  {author} {\bibinfo {author} {\bibfnamefont {N.}~\bibnamefont
  {Metropolis}}, \bibinfo {author} {\bibfnamefont {A.~W.}\ \bibnamefont
  {Rosenbluth}}, \bibinfo {author} {\bibfnamefont {M.~N.}\ \bibnamefont
  {Rosenbluth}}, \bibinfo {author} {\bibfnamefont {A.~H.}\ \bibnamefont
  {Teller}}, \ and\ \bibinfo {author} {\bibfnamefont {E.}~\bibnamefont
  {Teller}},\ }\href@noop {} {\bibfield  {journal} {\bibinfo  {journal} {J.
  Chem Phys}\ }\textbf {\bibinfo {volume} {21}},\ \bibinfo {pages} {1087}
  (\bibinfo {year} {1953})}\BibitemShut {NoStop}%
\bibitem [{\citenamefont {Pereira}\ \emph {et~al.}(2006)\citenamefont
  {Pereira}, \citenamefont {Guinea}, \citenamefont {Lopes~dos Santos},
  \citenamefont {Peres},\ and\ \citenamefont {Castro~Neto}}]{Pereira2006}%
  \BibitemOpen
  \bibfield  {author} {\bibinfo {author} {\bibfnamefont {V.~M.}\ \bibnamefont
  {Pereira}}, \bibinfo {author} {\bibfnamefont {F.}~\bibnamefont {Guinea}},
  \bibinfo {author} {\bibfnamefont {J.~M.~B.}\ \bibnamefont {Lopes~dos
  Santos}}, \bibinfo {author} {\bibfnamefont {N.~M.~R.}\ \bibnamefont {Peres}},
  \ and\ \bibinfo {author} {\bibfnamefont {A.~H.}\ \bibnamefont
  {Castro~Neto}},\ }\href@noop {} {\bibfield  {journal} {\bibinfo  {journal}
  {Phys. Rev. Lett.}\ }\textbf {\bibinfo {volume} {96}},\ \bibinfo {pages}
  {036801} (\bibinfo {year} {2006})}\BibitemShut {NoStop}%
\bibitem [{\citenamefont {Basko}(2008)}]{basko_resonant_2008}%
  \BibitemOpen
  \bibfield  {author} {\bibinfo {author} {\bibfnamefont {D.~M.}\ \bibnamefont
  {Basko}},\ }\href@noop {} {\bibfield  {journal} {\bibinfo  {journal} {Phys.
  Rev. B}\ }\textbf {\bibinfo {volume} {78}},\ \bibinfo {pages} {115432}
  (\bibinfo {year} {2008})}\BibitemShut {NoStop}%
\bibitem [{\citenamefont {Gargiulo}\ \emph {et~al.}(2014)\citenamefont
  {Gargiulo}, \citenamefont {Autès},\ and\ \citenamefont
  {Yazyev}}]{fernando_gabriel}%
  \BibitemOpen
  \bibfield  {author} {\bibinfo {author} {\bibfnamefont {F.}~\bibnamefont
  {Gargiulo}}, \bibinfo {author} {\bibfnamefont {G.}~\bibnamefont {Autès}}, \
  and\ \bibinfo {author} {\bibfnamefont {O.~V.}\ \bibnamefont {Yazyev}},\
  }\href@noop {} {\bibfield  {journal} {\bibinfo  {journal} {to be published}\
  } (\bibinfo {year} {2014})}\BibitemShut {NoStop}%
\bibitem [{\citenamefont {Büttiker}\ \emph {et~al.}(1985)\citenamefont
  {Büttiker}, \citenamefont {Imry}, \citenamefont {Landauer},\ and\
  \citenamefont {Pinhas}}]{buttiker_generalized_1985}%
  \BibitemOpen
  \bibfield  {author} {\bibinfo {author} {\bibfnamefont {M.}~\bibnamefont
  {Büttiker}}, \bibinfo {author} {\bibfnamefont {Y.}~\bibnamefont {Imry}},
  \bibinfo {author} {\bibfnamefont {R.}~\bibnamefont {Landauer}}, \ and\
  \bibinfo {author} {\bibfnamefont {S.}~\bibnamefont {Pinhas}},\ }\href@noop {}
  {\bibfield  {journal} {\bibinfo  {journal} {Phys. Rev. B}\ }\textbf {\bibinfo
  {volume} {31}},\ \bibinfo {pages} {6207} (\bibinfo {year}
  {1985})}\BibitemShut {NoStop}%
\bibitem [{\citenamefont {Choe}\ and\ \citenamefont
  {Chang}(2012)}]{choe_effect_2012}%
  \BibitemOpen
  \bibfield  {author} {\bibinfo {author} {\bibfnamefont {D.-H.}\ \bibnamefont
  {Choe}}\ and\ \bibinfo {author} {\bibfnamefont {K.~J.}\ \bibnamefont
  {Chang}},\ }\href@noop {} {\bibfield  {journal} {\bibinfo  {journal} {Nano
  Lett.}\ }\textbf {\bibinfo {volume} {12}},\ \bibinfo {pages} {5175} (\bibinfo
  {year} {2012})}\BibitemShut {NoStop}%
\bibitem [{\citenamefont {Uppstu}\ \emph {et~al.}(2014)\citenamefont {Uppstu},
  \citenamefont {Fan},\ and\ \citenamefont {Harju}}]{uppstu_obtaining_2014}%
  \BibitemOpen
  \bibfield  {author} {\bibinfo {author} {\bibfnamefont {A.}~\bibnamefont
  {Uppstu}}, \bibinfo {author} {\bibfnamefont {Z.}~\bibnamefont {Fan}}, \ and\
  \bibinfo {author} {\bibfnamefont {A.}~\bibnamefont {Harju}},\ }\href@noop {}
  {\bibfield  {journal} {\bibinfo  {journal} {Phys. Rev. B}\ }\textbf {\bibinfo
  {volume} {89}},\ \bibinfo {pages} {075420} (\bibinfo {year}
  {2014})}\BibitemShut {NoStop}%
\bibitem [{\citenamefont {Lee}\ and\ \citenamefont
  {Ramakrishnan}(1985)}]{lee_disordered_1985}%
  \BibitemOpen
  \bibfield  {author} {\bibinfo {author} {\bibfnamefont {P.~A.}\ \bibnamefont
  {Lee}}\ and\ \bibinfo {author} {\bibfnamefont {T.~V.}\ \bibnamefont
  {Ramakrishnan}},\ }\href@noop {} {\bibfield  {journal} {\bibinfo  {journal}
  {Rev. Mod. Phys.}\ }\textbf {\bibinfo {volume} {57}},\ \bibinfo {pages} {287}
  (\bibinfo {year} {1985})}\BibitemShut {NoStop}%
\bibitem [{\citenamefont {Weiße}\ \emph {et~al.}(2006)\citenamefont {Weiße},
  \citenamefont {Wellein}, \citenamefont {Alvermann},\ and\ \citenamefont
  {Fehske}}]{weise_kernel_2006}%
  \BibitemOpen
  \bibfield  {author} {\bibinfo {author} {\bibfnamefont {A.}~\bibnamefont
  {Weiße}}, \bibinfo {author} {\bibfnamefont {G.}~\bibnamefont {Wellein}},
  \bibinfo {author} {\bibfnamefont {A.}~\bibnamefont {Alvermann}}, \ and\
  \bibinfo {author} {\bibfnamefont {H.}~\bibnamefont {Fehske}},\ }\href@noop {}
  {\bibfield  {journal} {\bibinfo  {journal} {Rev. Mod. Phys.}\ }\textbf
  {\bibinfo {volume} {78}},\ \bibinfo {pages} {275} (\bibinfo {year}
  {2006})}\BibitemShut {NoStop}%
\end{thebibliography}%


\begin{thebibliography}{13}%
\makeatletter
\providecommand \@ifxundefined [1]{%
 \@ifx{#1\undefined}
}%
\providecommand \@ifnum [1]{%
 \ifnum #1\expandafter \@firstoftwo
 \else \expandafter \@secondoftwo
 \fi
}%
\providecommand \@ifx [1]{%
 \ifx #1\expandafter \@firstoftwo
 \else \expandafter \@secondoftwo
 \fi
}%
\providecommand \natexlab [1]{#1}%
\providecommand \enquote  [1]{``#1''}%
\providecommand \bibnamefont  [1]{#1}%
\providecommand \bibfnamefont [1]{#1}%
\providecommand \citenamefont [1]{#1}%
\providecommand \href@noop [0]{\@secondoftwo}%
\providecommand \href [0]{\begingroup \@sanitize@url \@href}%
\providecommand \@href[1]{\@@startlink{#1}\@@href}%
\providecommand \@@href[1]{\endgroup#1\@@endlink}%
\providecommand \@sanitize@url [0]{\catcode `\\12\catcode `\$12\catcode
  `\&12\catcode `\#12\catcode `\^12\catcode `\_12\catcode `\%12\relax}%
\providecommand \@@startlink[1]{}%
\providecommand \@@endlink[0]{}%
\providecommand \url  [0]{\begingroup\@sanitize@url \@url }%
\providecommand \@url [1]{\endgroup\@href {#1}{\urlprefix }}%
\providecommand \urlprefix  [0]{URL }%
\providecommand \Eprint [0]{\href }%
\providecommand \doibase [0]{http://dx.doi.org/}%
\providecommand \selectlanguage [0]{\@gobble}%
\providecommand \bibinfo  [0]{\@secondoftwo}%
\providecommand \bibfield  [0]{\@secondoftwo}%
\providecommand \translation [1]{[#1]}%
\providecommand \BibitemOpen [0]{}%
\providecommand \bibitemStop [0]{}%
\providecommand \bibitemNoStop [0]{.\EOS\space}%
\providecommand \EOS [0]{\spacefactor3000\relax}%
\providecommand \BibitemShut  [1]{\csname bibitem#1\endcsname}%
\let\auto@bib@innerbib\@empty
%</preamble>
\bibitem [{\citenamefont {Perdew}\ \emph {et~al.}(1996)\citenamefont {Perdew},
  \citenamefont {Burke},\ and\ \citenamefont
  {Ernzerhof}}]{s_perdew_generalized_1996}%
  \BibitemOpen
  \bibfield  {author} {\bibinfo {author} {\bibfnamefont {J.~P.}\ \bibnamefont
  {Perdew}}, \bibinfo {author} {\bibfnamefont {K.}~\bibnamefont {Burke}}, \
  and\ \bibinfo {author} {\bibfnamefont {M.}~\bibnamefont {Ernzerhof}},\
  }\href@noop {} {\bibfield  {journal} {\bibinfo  {journal} {Phys. Rev. Lett.}\
  }\textbf {\bibinfo {volume} {77}},\ \bibinfo {pages} {3865} (\bibinfo {year}
  {1996})}\BibitemShut {NoStop}%
\bibitem [{\citenamefont {Vanderbilt}(1990)}]{s_vanderbilt_soft_1990}%
  \BibitemOpen
  \bibfield  {author} {\bibinfo {author} {\bibfnamefont {D.}~\bibnamefont
  {Vanderbilt}},\ }\href@noop {} {\bibfield  {journal} {\bibinfo  {journal}
  {Phys. Rev. B}\ }\textbf {\bibinfo {volume} {41}},\ \bibinfo {pages} {7892}
  (\bibinfo {year} {1990})}\BibitemShut {NoStop}%
\bibitem [{\citenamefont {Monkhorst}\ and\ \citenamefont
  {Pack}(1976)}]{s_monkhorst_special_1976}%
  \BibitemOpen
  \bibfield  {author} {\bibinfo {author} {\bibfnamefont {H.~J.}\ \bibnamefont
  {Monkhorst}}\ and\ \bibinfo {author} {\bibfnamefont {J.~D.}\ \bibnamefont
  {Pack}},\ }\href@noop {} {\bibfield  {journal} {\bibinfo  {journal} {Phys.
  Rev. B}\ }\textbf {\bibinfo {volume} {13}},\ \bibinfo {pages} {5188}
  (\bibinfo {year} {1976})}\BibitemShut {NoStop}%
\bibitem [{\citenamefont {Giannozzi}\ \emph {et~al.}(2009)\citenamefont
  {Giannozzi}, \citenamefont {Baroni}, \citenamefont {Bonini}, \citenamefont
  {Calandra}, \citenamefont {Car}, \citenamefont {Cavazzoni}, \citenamefont
  {Ceresoli}, \citenamefont {Chiarotti}, \citenamefont {Cococcioni},
  \citenamefont {Dabo}, \citenamefont {Corso}, \citenamefont {Gironcoli},
  \citenamefont {Fabris}, \citenamefont {Fratesi}, \citenamefont {Gebauer},
  \citenamefont {Gerstmann}, \citenamefont {Gougoussis}, \citenamefont
  {Kokalj}, \citenamefont {Lazzeri}, \citenamefont {Martin-Samos},
  \citenamefont {Marzari}, \citenamefont {Mauri}, \citenamefont {Mazzarello},
  \citenamefont {Paolini}, \citenamefont {Pasquarello}, \citenamefont
  {Paulatto}, \citenamefont {Sbraccia}, \citenamefont {Scandolo}, \citenamefont
  {Sclauzero}, \citenamefont {Seitsonen}, \citenamefont {Smogunov},
  \citenamefont {Umari},\ and\ \citenamefont
  {Wentzcovitch}}]{s_giannozzi_quantum_2009}%
  \BibitemOpen
  \bibfield  {author} {\bibinfo {author} {\bibfnamefont {P.}~\bibnamefont
  {Giannozzi}}, \bibinfo {author} {\bibfnamefont {S.}~\bibnamefont {Baroni}},
  \bibinfo {author} {\bibfnamefont {N.}~\bibnamefont {Bonini}}, \bibinfo
  {author} {\bibfnamefont {M.}~\bibnamefont {Calandra}}, \bibinfo {author}
  {\bibfnamefont {R.}~\bibnamefont {Car}}, \bibinfo {author} {\bibfnamefont
  {C.}~\bibnamefont {Cavazzoni}}, \bibinfo {author} {\bibfnamefont
  {D.}~\bibnamefont {Ceresoli}}, \bibinfo {author} {\bibfnamefont {G.~L.}\
  \bibnamefont {Chiarotti}}, \bibinfo {author} {\bibfnamefont {M.}~\bibnamefont
  {Cococcioni}}, \bibinfo {author} {\bibfnamefont {I.}~\bibnamefont {Dabo}},
  \bibinfo {author} {\bibfnamefont {A.~D.}\ \bibnamefont {Corso}}, \bibinfo
  {author} {\bibfnamefont {S.~d.}\ \bibnamefont {Gironcoli}}, \bibinfo {author}
  {\bibfnamefont {S.}~\bibnamefont {Fabris}}, \bibinfo {author} {\bibfnamefont
  {G.}~\bibnamefont {Fratesi}}, \bibinfo {author} {\bibfnamefont
  {R.}~\bibnamefont {Gebauer}}, \bibinfo {author} {\bibfnamefont
  {U.}~\bibnamefont {Gerstmann}}, \bibinfo {author} {\bibfnamefont
  {C.}~\bibnamefont {Gougoussis}}, \bibinfo {author} {\bibfnamefont
  {A.}~\bibnamefont {Kokalj}}, \bibinfo {author} {\bibfnamefont
  {M.}~\bibnamefont {Lazzeri}}, \bibinfo {author} {\bibfnamefont
  {L.}~\bibnamefont {Martin-Samos}}, \bibinfo {author} {\bibfnamefont
  {N.}~\bibnamefont {Marzari}}, \bibinfo {author} {\bibfnamefont
  {F.}~\bibnamefont {Mauri}}, \bibinfo {author} {\bibfnamefont
  {R.}~\bibnamefont {Mazzarello}}, \bibinfo {author} {\bibfnamefont
  {S.}~\bibnamefont {Paolini}}, \bibinfo {author} {\bibfnamefont
  {A.}~\bibnamefont {Pasquarello}}, \bibinfo {author} {\bibfnamefont
  {L.}~\bibnamefont {Paulatto}}, \bibinfo {author} {\bibfnamefont
  {C.}~\bibnamefont {Sbraccia}}, \bibinfo {author} {\bibfnamefont
  {S.}~\bibnamefont {Scandolo}}, \bibinfo {author} {\bibfnamefont
  {G.}~\bibnamefont {Sclauzero}}, \bibinfo {author} {\bibfnamefont {A.~P.}\
  \bibnamefont {Seitsonen}}, \bibinfo {author} {\bibfnamefont {A.}~\bibnamefont
  {Smogunov}}, \bibinfo {author} {\bibfnamefont {P.}~\bibnamefont {Umari}}, \
  and\ \bibinfo {author} {\bibfnamefont {R.~M.}\ \bibnamefont {Wentzcovitch}},\
  }\href@noop {} {\bibfield  {journal} {\bibinfo  {journal} {J. Phys.: Condens.
  Matt.}\ }\textbf {\bibinfo {volume} {21}},\ \bibinfo {pages} {395502}
  (\bibinfo {year} {2009})}\BibitemShut {NoStop}%
\bibitem [{\citenamefont {Lin}\ \emph {et~al.}(2008)\citenamefont {Lin},
  \citenamefont {Ding},\ and\ \citenamefont {Yakobson}}]{s_lin_hydrogen_2008}%
  \BibitemOpen
  \bibfield  {author} {\bibinfo {author} {\bibfnamefont {Y.}~\bibnamefont
  {Lin}}, \bibinfo {author} {\bibfnamefont {F.}~\bibnamefont {Ding}}, \ and\
  \bibinfo {author} {\bibfnamefont {B.~I.}\ \bibnamefont {Yakobson}},\
  }\href@noop {} {\bibfield  {journal} {\bibinfo  {journal} {Phys. Rev. B}\
  }\textbf {\bibinfo {volume} {78}},\ \bibinfo {pages} {041402} (\bibinfo
  {year} {2008})}\BibitemShut {NoStop}%
\bibitem [{\citenamefont {Boukhvalov}\ \emph {et~al.}(2008)\citenamefont
  {Boukhvalov}, \citenamefont {Katsnelson},\ and\ \citenamefont
  {Lichtenstein}}]{s_boukhvalov_hydrogen_2008}%
  \BibitemOpen
  \bibfield  {author} {\bibinfo {author} {\bibfnamefont {D.~W.}\ \bibnamefont
  {Boukhvalov}}, \bibinfo {author} {\bibfnamefont {M.~I.}\ \bibnamefont
  {Katsnelson}}, \ and\ \bibinfo {author} {\bibfnamefont {A.~I.}\ \bibnamefont
  {Lichtenstein}},\ }\href@noop {} {\bibfield  {journal} {\bibinfo  {journal}
  {Phys. Rev. B}\ }\textbf {\bibinfo {volume} {77}},\ \bibinfo {pages} {035427}
  (\bibinfo {year} {2008})}\BibitemShut {NoStop}%
\bibitem [{\citenamefont {Büttiker}\ \emph {et~al.}(1985)\citenamefont
  {Büttiker}, \citenamefont {Imry}, \citenamefont {Landauer},\ and\
  \citenamefont {Pinhas}}]{s_buttiker_generalized_1985}%
  \BibitemOpen
  \bibfield  {author} {\bibinfo {author} {\bibfnamefont {M.}~\bibnamefont
  {Büttiker}}, \bibinfo {author} {\bibfnamefont {Y.}~\bibnamefont {Imry}},
  \bibinfo {author} {\bibfnamefont {R.}~\bibnamefont {Landauer}}, \ and\
  \bibinfo {author} {\bibfnamefont {S.}~\bibnamefont {Pinhas}},\ }\href@noop {}
  {\bibfield  {journal} {\bibinfo  {journal} {Phys. Rev. B}\ }\textbf {\bibinfo
  {volume} {31}},\ \bibinfo {pages} {6207} (\bibinfo {year}
  {1985})}\BibitemShut {NoStop}%
\bibitem [{\citenamefont {Mathon}\ and\ \citenamefont
  {Umerski}(2001)}]{s_mathon_theory_2001}%
  \BibitemOpen
  \bibfield  {author} {\bibinfo {author} {\bibfnamefont {J.}~\bibnamefont
  {Mathon}}\ and\ \bibinfo {author} {\bibfnamefont {A.}~\bibnamefont
  {Umerski}},\ }\href@noop {} {\bibfield  {journal} {\bibinfo  {journal} {Phys.
  Rev. B}\ }\textbf {\bibinfo {volume} {63}},\ \bibinfo {pages} {220403}
  (\bibinfo {year} {2001})}\BibitemShut {NoStop}%
\bibitem [{\citenamefont {Umerski}(1997)}]{s_umerski_closed-form_1997}%
  \BibitemOpen
  \bibfield  {author} {\bibinfo {author} {\bibfnamefont {A.}~\bibnamefont
  {Umerski}},\ }\href@noop {} {\bibfield  {journal} {\bibinfo  {journal} {Phys.
  Rev. B}\ }\textbf {\bibinfo {volume} {55}},\ \bibinfo {pages} {5266}
  (\bibinfo {year} {1997})}\BibitemShut {NoStop}%
\bibitem [{\citenamefont {Sanvito}\ \emph {et~al.}(1999)\citenamefont
  {Sanvito}, \citenamefont {Lambert}, \citenamefont {Jefferson},\ and\
  \citenamefont {Bratkovsky}}]{s_sanvito_general_1999}%
  \BibitemOpen
  \bibfield  {author} {\bibinfo {author} {\bibfnamefont {S.}~\bibnamefont
  {Sanvito}}, \bibinfo {author} {\bibfnamefont {C.~J.}\ \bibnamefont
  {Lambert}}, \bibinfo {author} {\bibfnamefont {J.~H.}\ \bibnamefont
  {Jefferson}}, \ and\ \bibinfo {author} {\bibfnamefont {A.~M.}\ \bibnamefont
  {Bratkovsky}},\ }\href@noop {} {\bibfield  {journal} {\bibinfo  {journal}
  {Phys. Rev. B}\ }\textbf {\bibinfo {volume} {59}},\ \bibinfo {pages} {11936}
  (\bibinfo {year} {1999})}\BibitemShut {NoStop}%
\bibitem [{\citenamefont {Weiße}\ \emph {et~al.}(2006)\citenamefont {Weiße},
  \citenamefont {Wellein}, \citenamefont {Alvermann},\ and\ \citenamefont
  {Fehske}}]{s_weise_kernel_2006}%
  \BibitemOpen
  \bibfield  {author} {\bibinfo {author} {\bibfnamefont {A.}~\bibnamefont
  {Weiße}}, \bibinfo {author} {\bibfnamefont {G.}~\bibnamefont {Wellein}},
  \bibinfo {author} {\bibfnamefont {A.}~\bibnamefont {Alvermann}}, \ and\
  \bibinfo {author} {\bibfnamefont {H.}~\bibnamefont {Fehske}},\ }\href@noop {}
  {\bibfield  {journal} {\bibinfo  {journal} {Rev. Mod. Phys.}\ }\textbf
  {\bibinfo {volume} {78}},\ \bibinfo {pages} {275} (\bibinfo {year}
  {2006})}\BibitemShut {NoStop}%
\bibitem [{\citenamefont {Tomczak}\ and\ \citenamefont
  {Biermann}(2009)}]{s_Tomczak09}%
  \BibitemOpen
  \bibfield  {author} {\bibinfo {author} {\bibfnamefont {J.~M.}\ \bibnamefont
  {Tomczak}}\ and\ \bibinfo {author} {\bibfnamefont {S.}~\bibnamefont
  {Biermann}},\ }\href@noop {} {\bibfield  {journal} {\bibinfo  {journal}
  {Phys. Rev. B}\ }\textbf {\bibinfo {volume} {80}},\ \bibinfo {pages} {085117}
  (\bibinfo {year} {2009})}\BibitemShut {NoStop}%
\bibitem [{\citenamefont {Roche}(1999)}]{s_roche_quantum_1999}%
  \BibitemOpen
  \bibfield  {author} {\bibinfo {author} {\bibfnamefont {S.}~\bibnamefont
  {Roche}},\ }\href@noop {} {\bibfield  {journal} {\bibinfo  {journal} {Phys.
  Rev. B}\ }\textbf {\bibinfo {volume} {59}},\ \bibinfo {pages} {2284}
  (\bibinfo {year} {1999})}\BibitemShut {NoStop}%
\end{thebibliography}%
\end{document}